\documentclass[a4paper,10pt]{article}
\usepackage{amsfonts}
\usepackage{amsmath}
\usepackage{amssymb}
\usepackage{mathrsfs}
\usepackage{subfigure}
\usepackage{graphicx}
\usepackage{epstopdf}
\usepackage{float}
\usepackage{color}
\usepackage{bm}
\usepackage{ulem}
\usepackage{xcolor}
\usepackage{appendix}
\usepackage{booktabs}
\usepackage{hyperref}
\usepackage{array}
\usepackage{cancel}
\usepackage{footnote}
\usepackage{authblk}  
\usepackage{geometry} 
\usepackage{cite}

\allowdisplaybreaks[4]

\hypersetup{colorlinks=true, linkcolor=blue, filecolor=blue, urlcolor=blue, citecolor=blue}

\begin{document}

\title{\textbf{Time Delay of Light in the Gravitational lensing of Supermassive Black Holes in Dark Matter Halos}}
	
\author{Chen-Kai Qiao \thanks{Email: chenkaiqiao@cqut.edu.cn}} 
\author{Ping Su \thanks{Email: suping@cqut.edu.cn}}
\affil{College of Science, Chongqing University of Technology, Banan, Chongqing, 400054, China}

\maketitle

\begin{abstract}
The dark matter halo has non-negligible effects on the gravitational lensing of supermassive black hole in the galaxy center. Our work presents a study on the time-delay of light in gravitational lensing of black holes enclosed by dark matter halos. To provide a precise description on the distribution of dark matter in galaxies, we choose several famous phenomenological dark matter halo models in astrophysics, including the NFW, Beta, Burkert and Moore models, to carry out the present study. Through numerically calculating the time-delay of light in gravitational lensing, a comparative analysis of the dark matter effects within different halo models has been performed. Assuming typical length scales associated with the galactic gravitational lensing, numerical results indicate that the NFW, Beta, Burkert and Moore dark matter halos can significantly enhance the time delay of light in gravitational lenisng of central supermassive black holes. The enhancing effect becomes more pronounced with a small dark matter halo scale and an increasing dark matter halo mass.
	
\ \ 

\end{abstract}

\section{Introduction \label{sec:1}}

The dark matter and dark energy have garnered significant attention and have stimulated a great number of extensive studies in the past years. The astrophysical evidences for dark matter have been witnessed in the observations of galactic rotation curves, \cite{Rubin1970,Corbelli2000}, bullet clusters \cite{Clowe2006}, large scale structure formation of the universe \cite{Davis1985}, cosmic microwave background and baryon acoustic oscillations \cite{WMAP2011,Planck2014}. These observations can hardly get well-pleasing explanations without assuming huge amount of dark matter in our universe. Particularly, the cosmic microwave background (CMB) observations suggested that $26.8\%$ of our universe is made up of dark patter, and $68.3\%$ of our universe is composed of dark energy \cite{Planck2014}. The most favorable candidates for dark matter are some unknown particles predicted by theories beyond the Standard Model, such as weakly interacting massive particles (WIMPs), axions, sterile neutrinos, etc \cite{Bertone2005,Boehm2004,Feng2009,Graham2015,Schumann2019,Boyarsky2019}. These unknown dark matters usually form the halo structures in large number of galaxies \cite{Cooray2002,Wang2019}, which could have notable impacts on the supermassive black holes in galaxies. Consequently, investigating the dark matter halo effects on supermassive black holes in the galaxy centers is of great significance, which provide us an approach to reveal the properties of dark matter and the its interplay with central black holes. Recently, a number of works to study the dark matter effects on black holes have been witnessed from various aspects, such as the circular geodesics \cite{Das2021,Rayimbaev2021,Konoplya2022}, black hole shadow \cite{HouX2018a,HouX2018b,Konoplya2019,Jusufi2019,Jusufi2020,Vagnozzi2022,Saurabh2021,Das2022,MaSJ2022,Pantig2022b,Ghosh2023,Yang2023,Wu2024,Macedo2024,Faraji2024}, accretion disk \cite{Fard2022}, chaos and thermodynamics \cite{XuZ2019,TaoJ2021,TaoJ2022,Ndongmo2023}, quasi-normal mode \cite{Jusufi2020b,LiuD2022,LiuD2024}, black hole echoes \cite{LiuD2021} and binary black hole merging process \cite{Bamber2022,Karydas2024}.  

Among the various aspects attempts explored in investigations of dark matter halos and black holes, gravitational lensing emerges as a highly active and powerful approach to study the dark matter and other matter fields. The weak and strong gravitational lensing have become significant important tools to test the astronomical and cosmological predictions \cite{Wambsganss1998,Bartelmann2001,Frittelli1998,Virbhadra1998,Virbhadra2000,Bozza2001,Virbhadra2002,Eiroa2002,Perlick2004,Virbhadra2009,Gibbons2008,Kogan2008,Bozza2008,Ishihara2016,Ishihara2016b,Crisnejo2018,Takizawa2020,Takizawa2020b,Li2020a,Li2020b,Huang2022,Tsukamoto2021,Tsukamoto2022,Asada2021,Asada2023,Perlick2022}, especially for dark matter halos in galaxies and galaxy clusters \cite{Clowe2006,Uitert2012,Brimioulle2013}. In the past five years, the dark matter halo effects on central black holes in the gravitational lensing were extensively analyzed using both pure theoretical models and astrophysical phenomenological dark matter models \cite{Haroon2019,Islam2020,Pantig2020,Pantig2021,Pantig2022,Pantig2022c,Atamurotov2021,Atamurotov2022,Qiao2023a,Qiao2023b}. The findings from these studies predict that the presence of dark matter halos can have non-negligible influences on the gravitational deflection angle of lights and the Einstein ring in gravitational lensing observations. However, in addition to the distortion effects in the light propagation (which results in gravitational deflection angle and Einstein ring), the gravitational lensing has another significant impacts on light traveling --- the time delay effect \cite{Shapiro1964,Virbhadra2008,Demorest2010,Christian2015,Papallo2015,Treu2016,Baker2017,Hou2018,Izmailov2019,Ng2020,Possel2020,Okcu2021,Dyadina2021,Lee2021,Azar2023,Malta2004}. Therefore, it is very useful to investigate the time delay of light in gravitational lensing to find whether the dark matter halo could have such notable influences on time delay compared with those in light deflection cases. 

In this work, the time delay of light in the gravitational lensing of supermassive black hole in the galaxy center is calculated and discussed, taking into account the presence of dark matter halos. To effectively highlight the effects of dark matter halos near the supermassive black hole in galaxy center, we adopt several astrophysical phenomenological dark matter halo model -- the Navarro-Frenk-White (NFW) model, isothermal Beta model, cusped Burkert model, and Moore model. These models have been widely used in the numerical simulations of galaxies and they have been successfully fitted by observational data in a large number of galaxies, which made them capable to give a precise description on dark matter distributions in galaxies \cite{Cavaliere1976, Burkert1995, Navarro1994, Navarro1995, Navarro1996, Moore1998, Moore1999, Salucci2000, Graham2005, Graham2006, Navarro2008, Brownstein2009, Dutton2014, Sofue2020}.  In this work, we use effective spacetime metrics to duel with the gravitational field of supermassive black holes combined with dark matter halos in galaxies. This treatment enables us to investigate the time delay of light in a simpler manner, without resorting to any complicated numerical calculations and techniques in the astrophysical gravitational lensing. The time delay of light are obtained by numerically solving the null geodesics. Specifically, to closely connect with the astrophysical gravitational lensing observations in typical galaxies (such as the Milky Way Galaxy), we select the dark matter halo scale $r_{\text{halo}} \sim $ 10 kpc and the supermassive black hole in the galaxy centers $M \sim 10^{7} M_{\odot}$. This choice of parameters could provide accurate predictions on the magnitude of time delay in most  astrophysical gravitational lensing observations of supermassive black holes in galaxy centers surrounded by dark matter halos. Furthermore, the time delay of light in the strong gravitational field cases, which is of great significance for theoretical studies, is also discussed in this work.

This paper is organized in the following way. The section \ref{sec:1} provides motivations and background introductions of this work. In section \ref{sec:2}, the effective spacetime metric generated by supermassive black holes in dark matter halos (described by NFW, Beta, Burkert, Moore models) is introduced. In section \ref{sec:3}, we describe the theoretical framework in this work. Results and discussions on the time delay of black holes surrounded by dark matter halo are presented in section \ref{sec:4}. Additionally, the analytical expressions on time delay derived in the weak gravitational field limit are given in Appendix \ref{appendix1}. The Appendix \ref{appendix2} gives calculations on the gravitational deflection angle of light. In this work, the natural unit $G=c=1$ is adopted.

\section{The Effective Spacetime Metric Generated by Supermassive Black Holes in Dark Matter Halos \label{sec:2}} 

In this section, we provide a concise derivation of the effective spacetime metric generated by black holes surrounded by dark matter halos in galaxy centers. For most galaxies, the dark matter halo can be approximately described by a spherically symmetric distribution. There are a number of astrophysical halo models that are commonly used to describe the dark matter distributions in our galaxies and other spiral galaxies \cite{Einasto1965, Cavaliere1976, Burkert1995, Navarro1994, Navarro1995, Navarro1996, Moore1998, Moore1999, Salucci2000, Graham2005, Graham2006, Navarro2008, Brownstein2009, Dutton2014, Sofue2020, Popolo2021, Shen2023,LiuD2023}. The famous phenomenological dark matter halo models include the NFW, Einasto, Beta, Burkert, Brownstein and Moore models \cite{Einasto1965, Cavaliere1976, Burkert1995, Navarro1994, Navarro1995, Navarro1996, Moore1998, Moore1999, Brownstein2009}. For simplicity, in the present work, we shall focus on the non-rotating black holes surrounded by dark matter halos whose spacetime metric can be expressed by, 
\begin{equation}
	\text{d\textit{s}}^2=-f(r) \text{dt}^2 +\frac{1}{f(r)}\text{dr}^2 +r^2\text{d$\theta $}^2+r^2\mathcal{\text{sin}}^2\theta \text{d$\phi $}^2.
\end{equation}
The effects of dark matter halo in gravitational lensing are through its mass profile and mass density. The dark matter halo mass profile is defined as
\begin{equation}
	M_{\text{DM}}(r) =4 \pi \int_{0}^{r} \mathcal{\rho}(r') r'^2 dr',
\end{equation}
where ${\rho}(r')$ is the density of dark matter distributions, whose detailed expressions are given by astrophysical phenomenological halo models. On one hand, from the Newtonian gravity, the tangential rotational velocity of a test particle or a massive celestial body in dark matter halo can be approximately calculated by ${v_{tg}}^2(r) \approx M_{\text{DM}}(r)/r$ using the mass profile of dark matter halo. Alternatively, for a test particle / massive celestial body moving in the equatorial plane of spherical symmetric spacetime, its rotation velocity can also be determined by the spacetime metric function $f(r)$ via equation \cite{Matos2000,Xu2018}
\begin{align}
	{v_{tg}}^2(r)=\frac{r}{\sqrt{f (r)}}\cdot  \frac{d \sqrt{f (r)}}{dr}=\frac{r (\text{dln} \sqrt{{f} (r)})}{dr}.\label{0.1}
\end{align}
Combining the rotational velocity in equation ({\ref{0.1}}) and its approximate expression ${v_{tg}}^2(r) \approx M_{\text{DM}}(r)/r$ in Newtonian gravity, the metric function generated by dark matter halo can be derived by solving the ordinary differential equation  (\ref{0.1})
\begin{align}
	f_{\text{DM}}(r) \approx \mathrm{exp}[2\int_{}^{} \frac{{v_{tg}}^2(r)}{r} \mathrm{d}r].\label{0.2}
\end{align}

In this work, we use several famous spherically symmetric dark matter profiles (the NFW, Beta, Burkert, Moore models). These models were proposed and developed for many years. In particular, Cavaliere and Fusco-Femiano proposed the isothermal Beta model in 1976 when studying the distribution of matter in galaxies \cite{Cavaliere1976}. In 1990s, J. F. Navarro, C. S. Frenk, S. D. M. White proposed a universal density profile relation for dark matter during the large scale simulations \cite{Navarro1995, Navarro1996}, this renowed model successfully reproduced the observed rotational curves in a large number of galaxies and it is named as the NFW halo model. Meanwhile, A. Burkert proposed an empirical density profile that successfully fitted the rotation curves of four dark matter-dominated dwarf galaxies in 1995 \cite{Burkert1995, Salucci2000}. In 1999, B. Moore et al. suggested that the dark matter density profile could have a cusp proportional to $r^{-1.5}$ in both galaxy-sized and cluster-sized halos \cite{Moore1999}. These dark matter profiles have been widely used in theoretical predictions and numerical simulations in physics and astronomy. The dark matter distributions of the NFW, Beta, Burkert and Moore models can be expressed as
\begin{subequations}
	\begin{eqnarray}
		&& \rho_{\text{NFW}}(x) = \frac{\rho_{0}}{x(1+x^{2})},
		\\
		&& \rho_{\text{Beta}}(x) = \frac{\rho_{0}}{(1+x^{2})^{3/2}},
		\\
		&& \rho_{\text{Burkert}}(x) = \frac{\rho_{0}}{(1+x)(1+x^2)},
	    \\
		&& \rho_{\text{Moore}}(x) = \frac{\rho_{0}}{x^{3/2}(1+x^{3/2})},
	\end{eqnarray}
\end{subequations}
We have used the variable $x=r/r_{\text{halo}}$ to express the dark matter density, with $\rho_{0}$ and $r_{\text{halo}}$ to be the characteristic density and radius of dark matter halos respectively. Among these dark matter distribution, the isothermal Beta model and Burkert model are cored halo models with a smooth density profile near the galaxy center, while the NFW and Moore models are cusp halo models with a rapidly increased density profile near the galaxy center.

To get the effective spacetime metric of black holes surrounded by above dark matter halos, one can combine the dark matter density profile $\rho$ and the central black hole mass $M$ to the energy-momentum tensor $T_{\mu\nu} $ in the Einstein field equation  \cite{Xu2018}. It turns out that the effective spacetime metric can be decomposed into two parts $f(r)=f_{\text{DM}}(r)-\frac{2M}{r}$, where $f_{\text{DM}}(r)$ in equation (\ref{0.2}) describes the dark matter halo and $-\frac{2M}{r}$ describes the effects of supermassive black hole in the galaxy center \cite{Xu2018}. Eventually, the metric function of the supermassive black holes in NFW, Beta, Burkert and Moore dark matter halos becomes:
\begin{subequations}
	\begin{eqnarray}
		&& f_{\text{NFW}}(r) = (1+x)^{ -\frac{8 \pi k_{\text{DM}}}{r} } - \frac{2M}{r}, 
		\label{NFW}
		\\
		&& f_{\text{Beta}}(r)= e^{-\frac{8 \pi k_{\text{DM}}}{r} \sinh^{-1} x} - \frac{2M}{r}
		\label{Beta} 
		\\
		&& f_{\text{Burkert}}(r) = e^{\frac{4 \pi  k_{\text{DM}} }{r}(1+x) \arctan x}
		\cdot (1+x)^{-\frac{4 \pi  k_{\text{DM}} }{r}(1+x)} 
		\cdot (1+x^2)^{\frac{2 \pi k_{\text{DM}}}{r} (x-1)} 
		- \frac{2M}{r},
		\label{Burkert}
		\\
		&& f_{\text{Moore}}(r) = e^{\frac{16 \pi k_{\text{DM}}}{\sqrt{3} r_{\text{halo}}} \arctan\frac{2 \sqrt{x}-1}{\sqrt{3}}}
		\cdot (1+x^{3/2})^{-\frac{16 \pi k_{\text{DM}}}{3r}}
		\cdot \bigg( \frac{1+x-\sqrt{x}}{1+x+2\sqrt{x}} \bigg)^{-\frac{8 \pi k}{3r_{\text{halo}}}} 
		- \frac{2M}{r},
		\label{Moore}
	\end{eqnarray}
\end{subequations}
where $M$ is the mass of the supermassive black hole, and $k_{\text{DM}} = \rho_{0} \cdot r_{\text{halo}}^3$ can be used to give an estimation of the dark matter halo mass. In a typical galaxy, the sale of dark matter halo is much larger than the horizon of central supermassive black hole, $r_{\text{H}} \ll  r_{\text{halo}}$. Furthermore, for the majority of the astrophysical observed gravitational lensing of supermassive black hole in the galaxy center, the minimal radius $r_{0}$ in the photon orbit is usually much smaller than the dark matter halo scale length. Therefore, in the numerical calculations, it is reasonable to establish the following relation 
\begin{equation}
	r_{\text{H}} \ll r_{0} \sim b \ll r_{\text{halo}}.
\end{equation}
Here, $r_{H}$ is the horizon radius of central supermassive black hole, $r_{\text{halo}}$ denotes the scale of dark matter halo, and $b$ is the impact parameter for light rays in gravitational lensing observations
\footnote{
Among the four aforementioned dark matter models, the effective metrics for NFW and Beta models in equation (\ref{NFW}) and (\ref{Beta}) are asymptotically flat, while the Burkert and Moore models' effective metrics given in equation (\ref{Burkert}) and (\ref{Moore}) are not asymptotically flat in $r \rightarrow \infty$ limit. In order to consistent with the description, we must manually choose a maximum cutoff value for radius variable $r_{\text{max}}$ such that the effective metrics for Burkert and Moore models fall to describe the spacetime in region $r>r_{\text{max}}$. However, an astrophysical galaxy in the presence of a dark matter halo is usually regraded as an isolated system, naturally there should be infinite distance observer in asymptotically flat region (if we neglect the cosmological constant effects). In this work, we simply set the effective metrics for Burkert and Moore models do not change at $r>r_{\text{max}}$, such that observers in $r>r_{\text{max}}$ can be effectively regarded as infinite distance observers in asymptotically flat region.
\begin{equation}
	f(r) = \bigg\{
	       \begin{aligned}
	       	  f_{\text{Burkert}}(r) \ \text{or} \  f_{\text{Moore}}(r) & & r < r_{\text{max}} \\
	       	  f_{\text{Burkert}}(r_{\text{max}}) \ \text{or} \  f_{\text{Moore}}(r_{\text{max}}) & & r \ge r_{\text{max}}
	       \end{aligned}
	       \nonumber
\end{equation} 
Particularly, in the calculation of time delay, an additional rescaled factor $f_{\text{rescale}} = \sqrt{1/f(r_{\text{max}})}$ should be used for Burkert and Moore dark matter models to connected with the infinite distance observer's time coordinate in asymptotically flat region (namely employing the transformation $t \rightarrow \tilde{t}$ with rescaled factor such that $f(r_{\text{max}})dt^{2} = d \tilde{t}^{2} \Leftrightarrow d\tilde{t} = \frac{dt}{ f_{\text{rescale}} }$ holds in asymptotically flat region). 
\label{footnote1} }.

\section{Theoretical Framework : Time Delay of Light \label{sec:3}}

This section provides an introduction on the theoretical framework of our work. The time-delay of light in a gravitational field can be calculated by solving the differential equations of null geodesics. Over the past few decades, this approach has been tested by large numbers of observations, and it has become a widely adopted approach in physics and astronomy, especially in the investigation of gravitational deflection and time delay \cite{Iyer2007,Tsukamoto2017,Kim2021,Lee2021b,Atamurotov2022a,Tsukamoto2022,Soares2023,Chowdhuri2023,Nazari2022,Chen2023,Nash2023}. 

For a spherically symmetric spacetime
\begin{equation}
	d\tau^{2} = f(r)dt^{2} -\frac{1}{f(r)}dr^{2} 
	-r^{2}(d\theta^{2}+\sin^{2}\theta d\phi^{2})
	\label{spacetime metric spherically symmetric}
\end{equation}
the following conserved quantities can be introduced 
\begin{subequations}
	\begin{eqnarray}
		J & \equiv & r^{2}\sin^{2}\theta \frac{d\phi}{d\lambda}
		\\
		E & \equiv &  f(r)\frac{dt}{d\lambda}
		\\
		\epsilon & \equiv & g_{\mu\nu}dx^{\mu}dx^{\nu} 
		= f(r) \bigg( \frac{dt}{d\lambda} \bigg)^2
		- \frac{1}{f(r)} \bigg( \frac{dr}{d\lambda} \bigg)^{2}
		- r^{2} \bigg( \frac{d\theta}{d\lambda} \bigg)^{2} 
		- r^{2}\sin^{2}\theta \bigg( \frac{d\phi}{d\lambda} \bigg)^{2}
	\end{eqnarray}
\end{subequations}
Here, $\lambda$ is an affine parameter, $J$ is the conserved angular momentum along a particle orbit, and $E^{2}/2$ can be viewed as the conserved energy along a particle orbit. For test particles moving in the equatorial plane $\theta=\pi/2$, the reduced differential equation can be obtained using these conserved quantities
\begin{equation}
		\frac{1}{2} \bigg( \frac{dr}{d\lambda} \bigg)^{2} + \frac{1}{2} f(r) \bigg[ \frac{J^{2}}{r^{2}} + \epsilon \bigg]
		= \frac{1}{2} \bigg( \frac{dr}{d\lambda} \bigg)^{2} + V_{\text{eff}}(r)
		= \frac{1}{2}E^{2} \label{reduced differential equation} 
\end{equation}
Here, $V_{\text{eff}}(r) = \frac{f(r)}{2} \big[ \frac{J^{2}}{r^{2}} + \epsilon \big]$ is the effective potential of test particles moving in the spherically symmetric gravitational field, and the impact parameter is defined as $b\equiv |J/E|$. For massless (or massive) particles traveling along null (or timelike) geodesics, the quantity $\epsilon$ takes the value $\epsilon=0$ (or $\epsilon=1$). Particularly, when focusing on a photon orbit, one can get the following relation from equation (\ref{reduced differential equation}) 
\begin{equation}
	\frac{dr}{dt} = \frac{dr}{d\lambda} \cdot \frac{d\lambda}{dt}
	= \pm f(r) \sqrt{1 - b^{2}\cdot\frac{f(r)}{r^{2}}}  
\end{equation}
where we have used $E = f(r)\frac{dt}{d\lambda}$, $\epsilon=0$ for massless photon, and the definition of impact parameter $b\equiv |J/E|$.
The plus and minus sign $\pm$ can be determined in the following way. When a particle begin to move along the scattering orbit from the source position $r_{\text{S}}$, the radial coordinate $r$ decreases with the time passes, until this particle reaches the closest distance $r=r_0$ to central supermassive black hole. After passing the turning point $r=r_{0}$, the radial coordinate of particle starts to increase as time progresses. Namely, we have the relations 
\begin{subequations}
	\begin{eqnarray}
		&& \frac{dr}{dt} = - f(r) \sqrt{1 - b^{2}\cdot\frac{f(r)}{r^{2}}} < 0 \ \ \ \text{photon moving from source position $r=r_{\text{S}}$ to the tuning point $r=r_{0}$} 
		\nonumber
		\\
		&& \frac{dr}{dt} = f(r) \sqrt{1 - b^{2}\cdot\frac{f(r)}{r^{2}}} > 0 \ \ \ \ \ \text{photon moving from the tuning point $r=r_{0}$ to observer position $r=r_{\text{O}}$}  
		\nonumber
	\end{eqnarray}
\end{subequations} 

In the gravitational lensing, when the light source and observer are located at $r=r_{\text{S}}$ and $r=r_{\text{O}}$ respectively, the time-delay of light during the propagation in the gravitational field can be expressed as 
\footnote{ 
This formula works well for NFW and Beta dark matter halo models with asymptotically flat effective metircs. However, for asymptotically non-flat effective metrics for Burkert and Moore models, an additional rescaled factor $f_{\text{rescale}} = 1/f(r_{\text{max}})$ should be included to connected with the infinite distance observer's time coordinate in asymptotically flat region, as we have explained in footnote \ref{footnote1}. In this way, the time delay of light measured by a distant observer at asymptotically flat region should become 
\begin{equation}
	\Delta T = \tilde{T} - T_{0} 
	= \int_{r_{0}}^{r_{\text{S}}} \frac{d\tilde{r}}{\tilde{f}(\tilde{r}) \sqrt{1-\frac{b^{2} \tilde{f}(\tilde{r})}{\tilde{r}^{2}}}}
	+ \int_{r_{0}}^{r_{\text{O}}} \frac{d\tilde{r}}{\tilde{f}(\tilde{r}) \sqrt{1-\frac{b^{2} \tilde{f}(\tilde{r})}{\tilde{r}^{2}}}}
	- \sqrt{r_{\text{S}}^{2}-r_{0}^{2}} - \sqrt{r_{\text{O}}^{2}-r_{0}^{2}}
	\nonumber
\end{equation} 
where we have used the rescaled factor to make the coordinate transformation $ d\tilde{t} = \frac{dt}{ f_{\text{rescale}} }$, $ d\tilde{r} = f_{\text{rescale}} \cdot dr$ and $\tilde{f}(\tilde{r}) = f_{\text{rescale}}^{2} \cdot f(r)$ such that $\tilde{f}(\tilde{r}) \cdot d\tilde{t}^{2} = f(r) \cdot dt^{2}$. In the present work, the $r_{\text{max}} = 100 \text{kpc}$ is assumed in the numerical calculations. The rescaled factors are very close to 1 for Burkert and Moore models ($1- f_{\text{rescale}} \sim 10^{-9} - 10^{-8}$), while it could still influence the time delay results because the radius $r_{\text{O}}$, $r_{\text{S}}$ could be very large in astrophysical galactic gravitational lensing observations. In the first version of our arXiv preprint, the rescaled factor is not considered. This is the reason that we got the incorrect negative results for time delay $\Delta T$ for Burkert model in the first version of the arXiv preprint.
\label{footnote2} }
\begin{eqnarray}
	\Delta T = T - T_{0} 
	& = & -\int_{r_{\text{S}}}^{r_{0}} \frac{dr}{f(r)\sqrt{1-\frac{b^{2} f(r)}{r^{2}}}}
	      + \int_{r_{0}}^{r_{\text{O}}} \frac{dr}{f(r)\sqrt{1-\frac{b^{2} f(r)}{r^{2}}}}
	      - T_{0} \nonumber
	      \\
	& = & \int_{r_{0}}^{r_{\text{S}}} \frac{dr}{f(r)\sqrt{1-\frac{b^{2} f(r)}{r^{2}}}}
          + \int_{r_{0}}^{r_{\text{O}}} \frac{dr}{f(r)\sqrt{1-\frac{b^{2} f(r)}{r^{2}}}}
          - \sqrt{r_{\text{S}}^{2}-r_{0}^{2}} - \sqrt{r_{\text{O}}^{2}-r_{0}^{2}} 
	\label{time delay for null geodesic}
\end{eqnarray}
with $T_{0} =\sqrt{r_{\text{S}}^{2}-r_{0}^{2}}+\sqrt{r_{\text{O}}^{2}-r_{0}^{2}}$ to be the time period during the light propagation process without the presence of gravitational field. In the expression (\ref{time delay for null geodesic}), it is obvious that the time-delay $\Delta T$ increases monotonically as the coordinates of light source and observer $r_{\text{S}}$, $r_{\text{O}}$ increase. In the integration process, the turning point $r=r_{0}$ must be solved. For the closet distance to central black hole, the derivative $dr/d\lambda$ in equation (\ref{reduced differential equation}) vanishes automatically
\begin{eqnarray}
	\frac{dr}{d\lambda}\bigg|_{r=r_{0}} = 0 
	\ & \Rightarrow & \
	\frac{1}{2}f(r_{0}) \frac{J^{2}}{r_{0}^{2}} = \frac{1}{2}E^{2} \nonumber
	\\
	\ & \Rightarrow & \ b^{2} = \frac{J^{2}}{E^{2}} = \frac{r_{0}^{2}}{f(r_{0})} \label{impactparameter-closetdistence}
\end{eqnarray}
Given the impact parameter $b$ for a photon orbit, the closet distance $r_{0}$ can be solved from this equation.

\section{Results and Discussions \label{sec:4}}

This section presents the numerical results on time-delays of light in the gravitational lensing of supermassive black holes surrounded by dark matter halos. Firstly, we present the time delay results calculated in the weak gravitational field cases in subsection \ref{sec:4a}. In this subsection, we adopt typical parameter values for galactic black hole and galactic dark matter halos, which are closely relevant to the astrophysical gravitational lensing observations in galaxies and galaxy centers. Secondly, we give a concise discussion on time delays in the strong gravitational field cases in subsection \ref{sec:4b}, which is extremely important to theoretical investigations of gravity theory.

\subsection{Time Delay in the Typical Galactic Gravitational Lensings (Weak Gravitational Field Cases) \label{sec:4a}}

This subsection presents the numerical results on time-delays of light for typical galactic gravitational lensings, where the supermassive black holes located in the galaxy center are surrounded by corresponding galactic dark matter halos. Most of these observations are measured in the weak gravitational cases, with a large distance from the light source and observer to the supermassive black hole.

\begin{figure}
	\centering
	\includegraphics[width=0.85\textwidth]{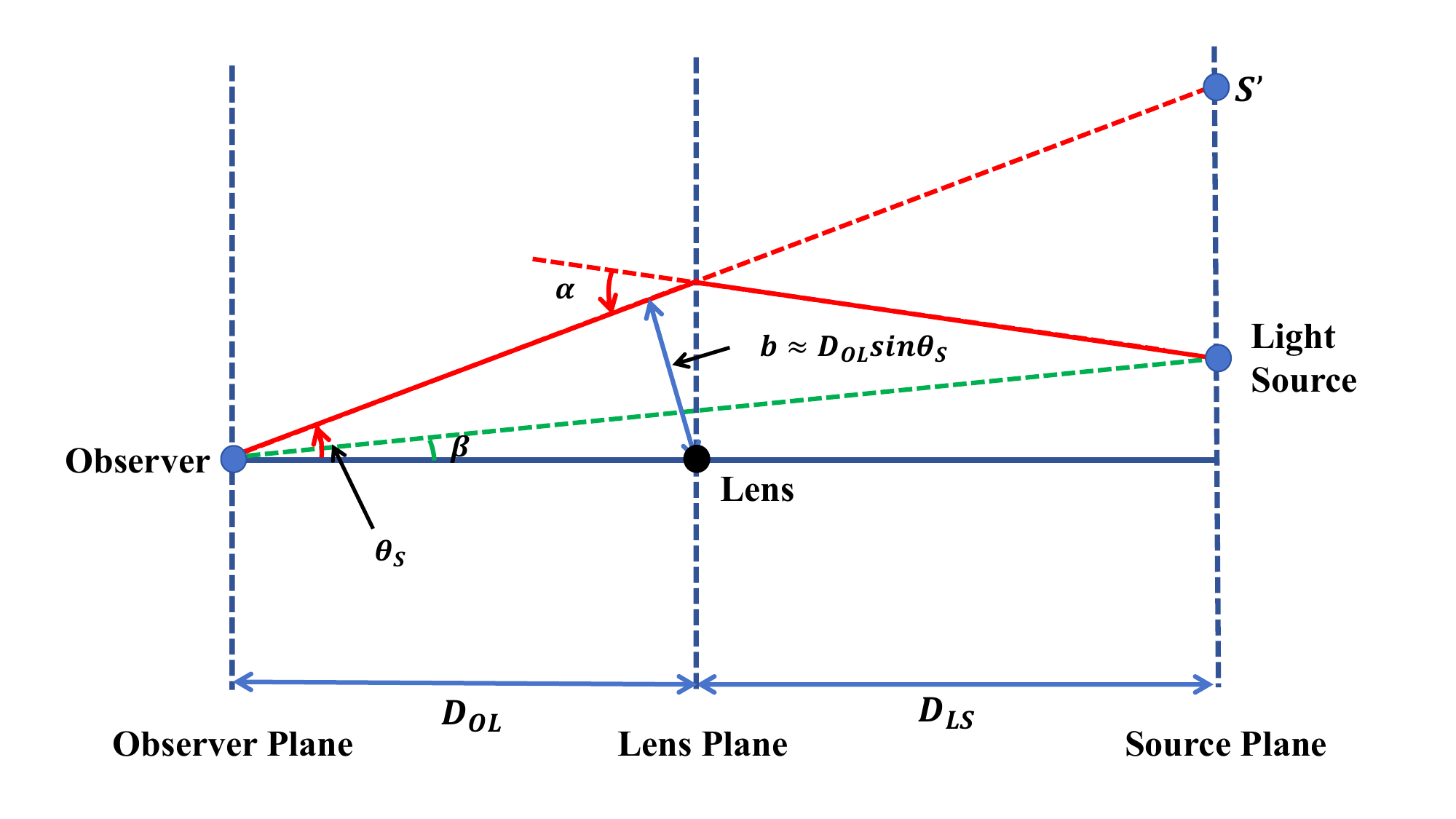}
	\caption{This figure illustrates the gravitational lensing of supermassive black hole in the weak deflection limit. In this figure, the positions of the observer, luminous light source, and the central supermassive black hole (acting as a gravitational lens) are labeled respectively. The gravitational deflection angle of light $\alpha$, angular position of the light source $\beta$, angular position of the lensed images $\theta_{S}$, and impact parameter $b$ have been shown in the figure. \label{figure1}}
\end{figure}

\begin{figure*}
	\includegraphics[width=0.5\textwidth]{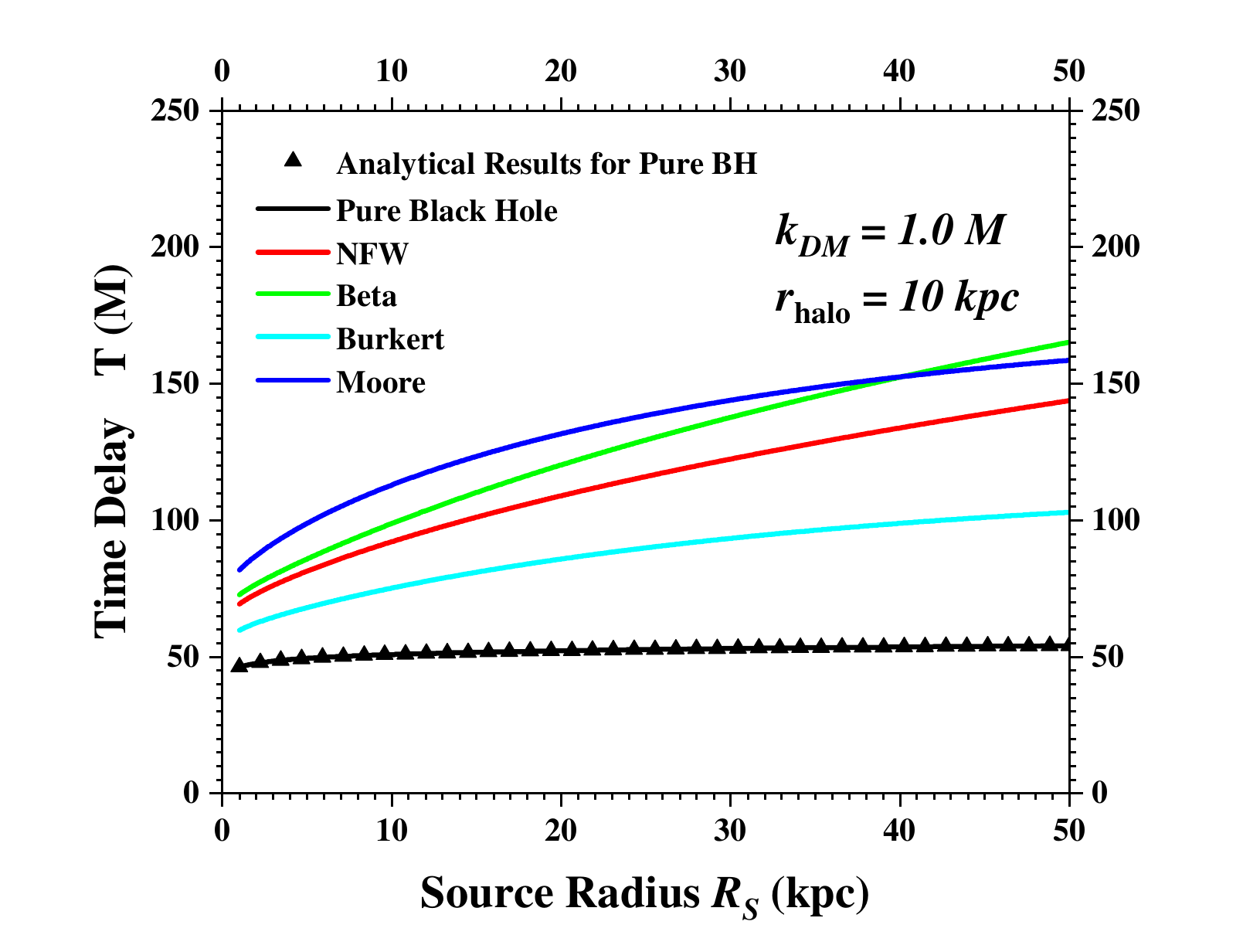}
	\includegraphics[width=0.5\textwidth]{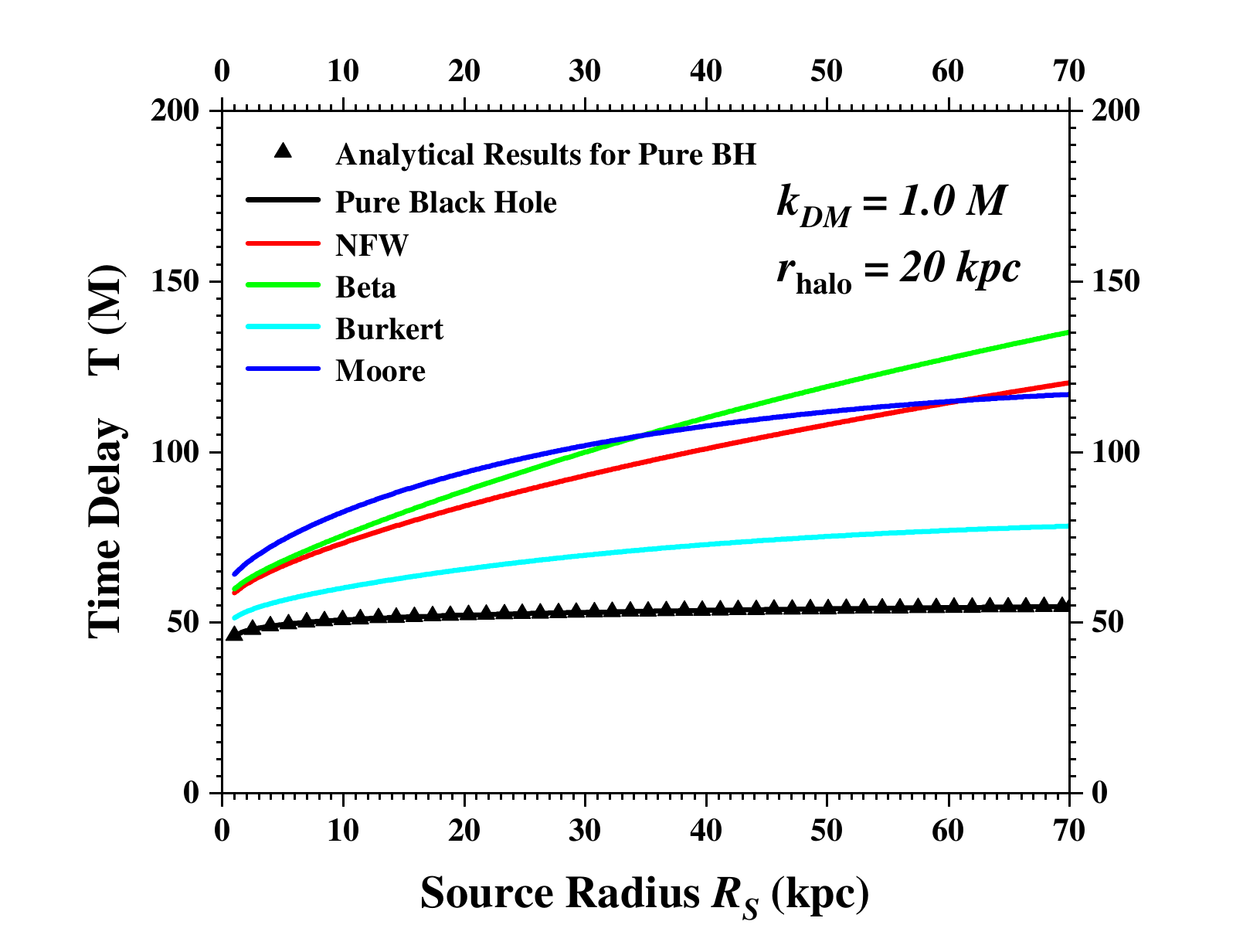}
	\includegraphics[width=0.5\textwidth]{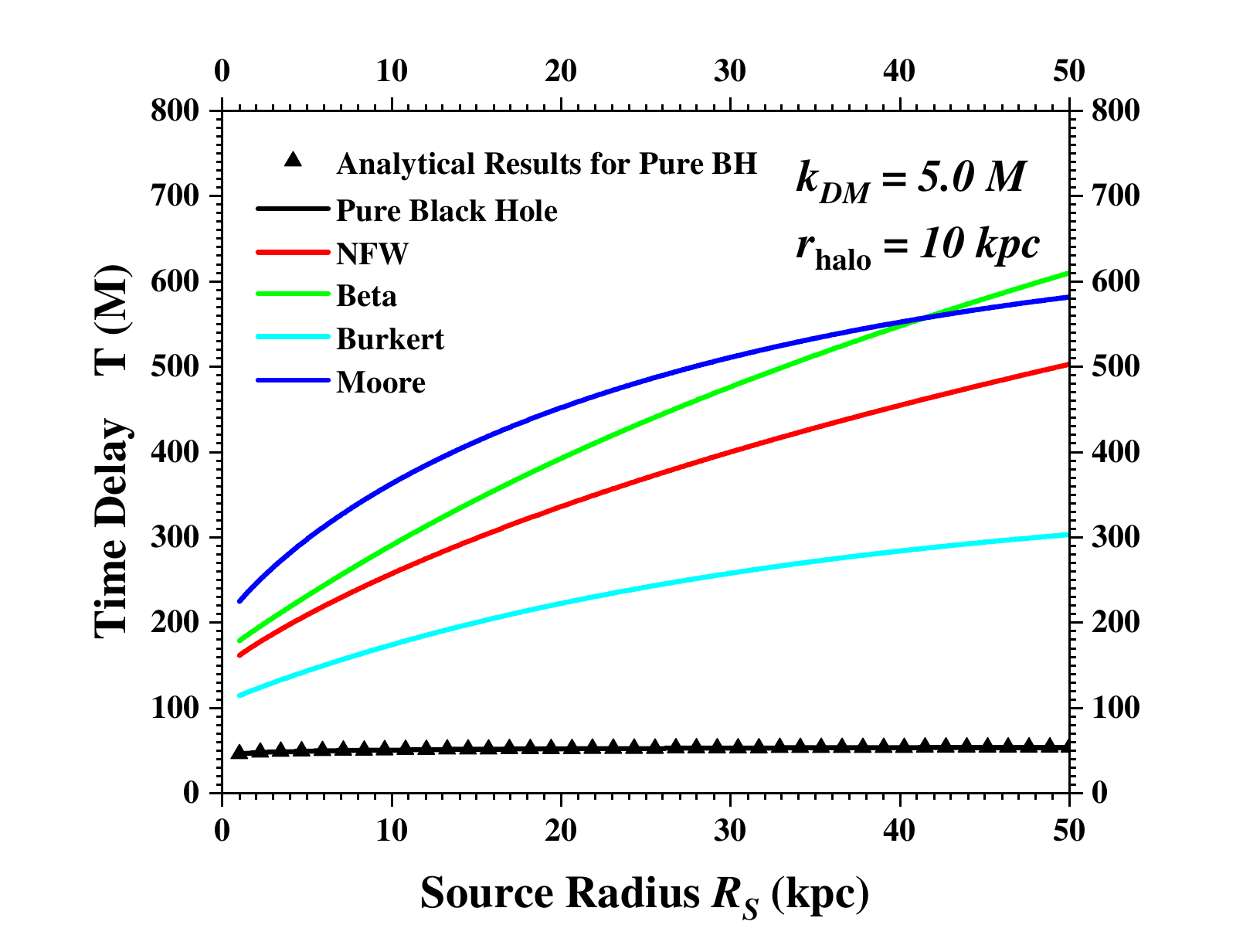}
	\includegraphics[width=0.5\textwidth]{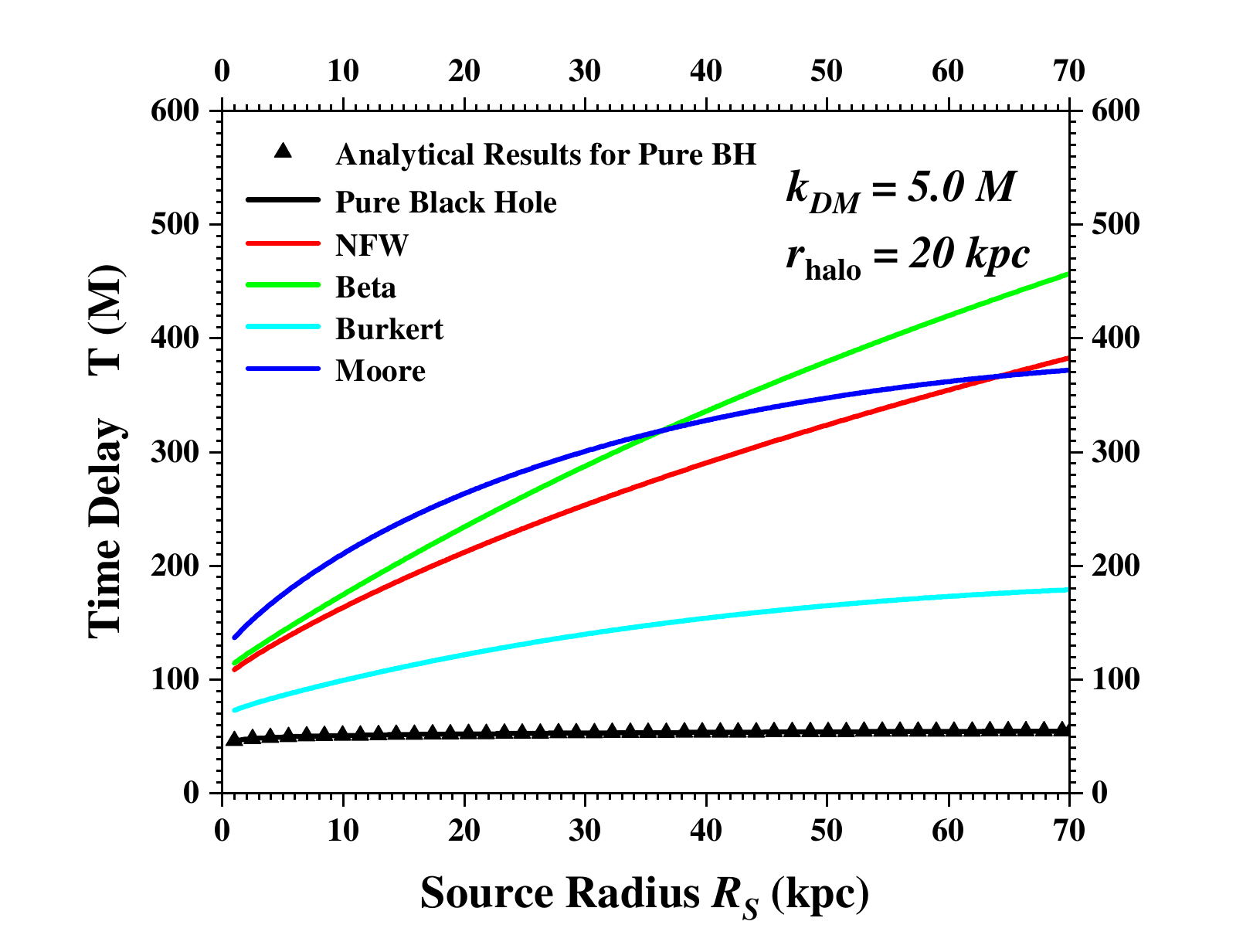}
	\includegraphics[width=0.5\textwidth]{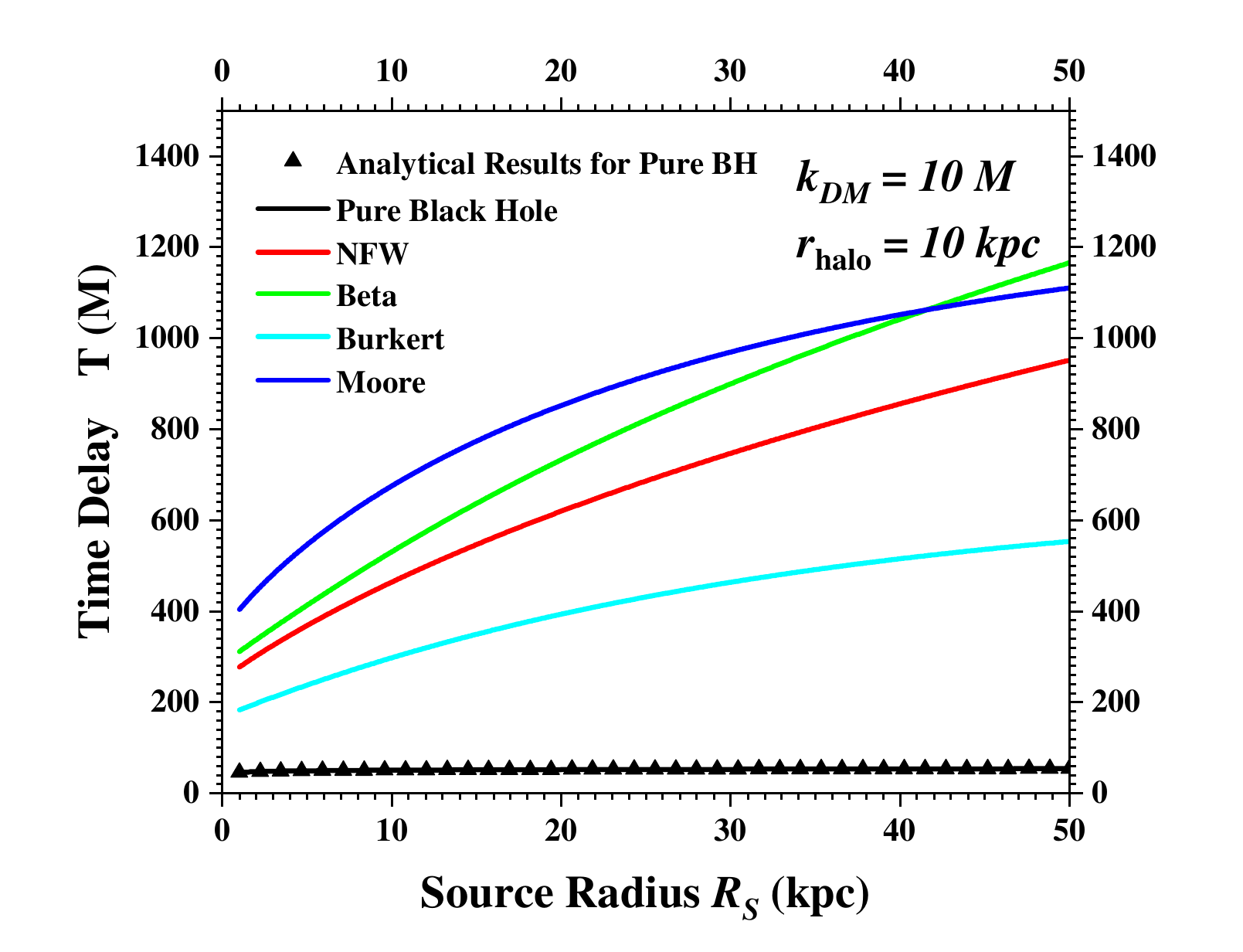}
	\includegraphics[width=0.5\textwidth]{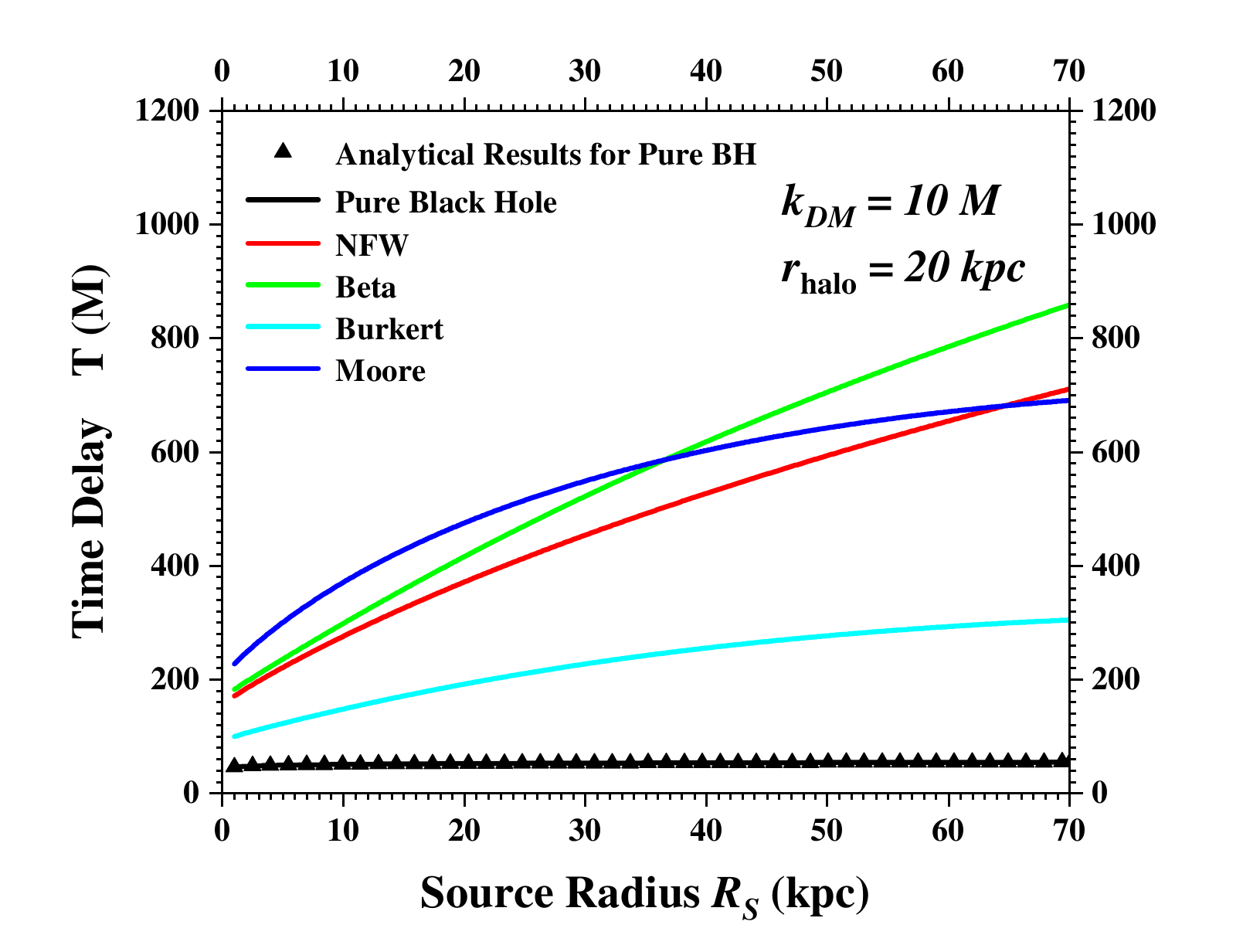}
	\caption{The time delay of light in gravitational lensing of black hole surrounded by dark matter halos. This figure presents the numerical results calculated using the geodesic approach in equation (\ref{time delay for null geodesic}) (obtained by solving the trajectories of null geodesics). The numerical results presented here are obtained within the NFW, Bata, Burkert, Moore dark matter halos models. The dark matter characteristic mass is set as $k_{\text{DM}}=M$, $k_{\text{DM}}=5M$ $k_{\text{DM}}=10M$, and the dark matter halo scale is selected to be $r_{\text{halo}}=10^{10} M \sim 10$ kpc and $r_{\text{halo}}=2 \times 10^{10} M \sim 20$ kpc in the left and right panels respectively. The time delay of light in gravitational lensing of pure black holes without the presence of dark matter halos is illustrated using black curves for comparisons. Additionally, the analytical results of Shapiro time delay $\Delta T = 2M\big(1+\ln(\frac{r_{\text{O}}r_{\text{S}}}{r_{0}^{2}})\big)$ for the cases of pure black hole in the weak field approximation are also depicted in the figure. In this figure, the location of observer is fixed as $r_{\text{O}}=10^{10} M \sim 10$ kpc, while the position of light source varies from $r_{\text{S}} = 1 \ \text{kpc}$ to $r_{\text{S}} =50 \ \text{kpc}$ in the left panel and from $r_{\text{S}} = 1 \ \text{kpc}$ to $r_{\text{S}} = 70 \ \text{kpc}$ in the right panel. The impact parameter in the photon orbit is $b=10^{5} M$ (in which the produced Einstein ring is roughly $\theta_{E} \sim$ arcsec).}
	\label{figure2}
\end{figure*}

To establish a closer connection with the astrophysical gravitational lensing observations in typical galaxies (such as the gravitational lensing of Sgr*A in Milky Way Galaxy), we select the parameters for supermassive black hole and the dark matter halo based on observed data from galaxy observations. For most spiral galaxies, the scale of the dark matter halo is approximately $r_{\text{halo}} \sim $ 10 kpc. The mass scale of supermassive black hole in the galaxy centers is roughly $M=10^{7} M_{\odot} \sim 10^{7} $ km $\sim 10^{-6}$ pc 
\footnote{For instance, the supermassive black hole Sgr*A in the Milky Way Galaxy has the mass $M = 4.3 \times 10^{6} M_{\odot}$ \label{footnote3}}, 
which made the relation $r_{\text{halo}} \sim 10^{10} M$ valid for spiral galaxies. The value of impact parameter $b$ in the photon orbit can be estimated through gravitational lensing observations using the observed magnitude of Einstein ring $\theta_{E}$. In the weak deflection limit, the impact parameter in photon orbit can be approximated as $b \sim D_{OL}  \sin\theta_{S} \sim D_{OL} \sin\theta_{E}$, as depicted in figure \ref{figure1}. In the context of gravitational lensing by supermassive black holes, the magnitude of Einstein ring is approximately $\theta_{E} \sim$ arcsec $\sim 10^{-5}$ rad, which implies the impact parameter to be $b \sim 0.1$ pc $\sim 10^{5} M$ (assuming the distances between observer, lensed supermassive black hole and light source satisfy $D_{\text{OL}} \sim D_{\text{LS}} \sim 10$ kpc). 

\begin{figure*}
	\includegraphics[width=0.5\textwidth]{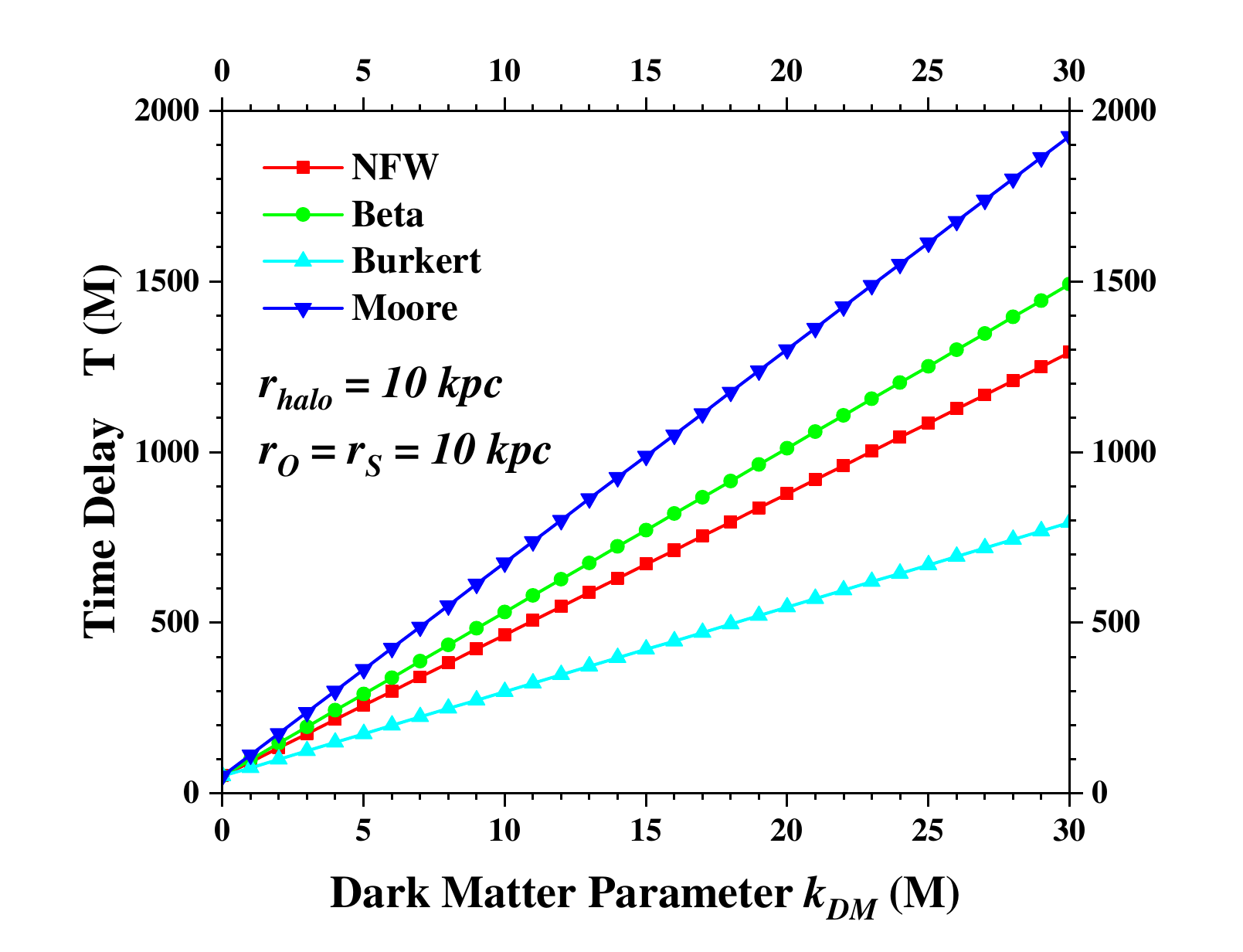}
	\includegraphics[width=0.5\textwidth]{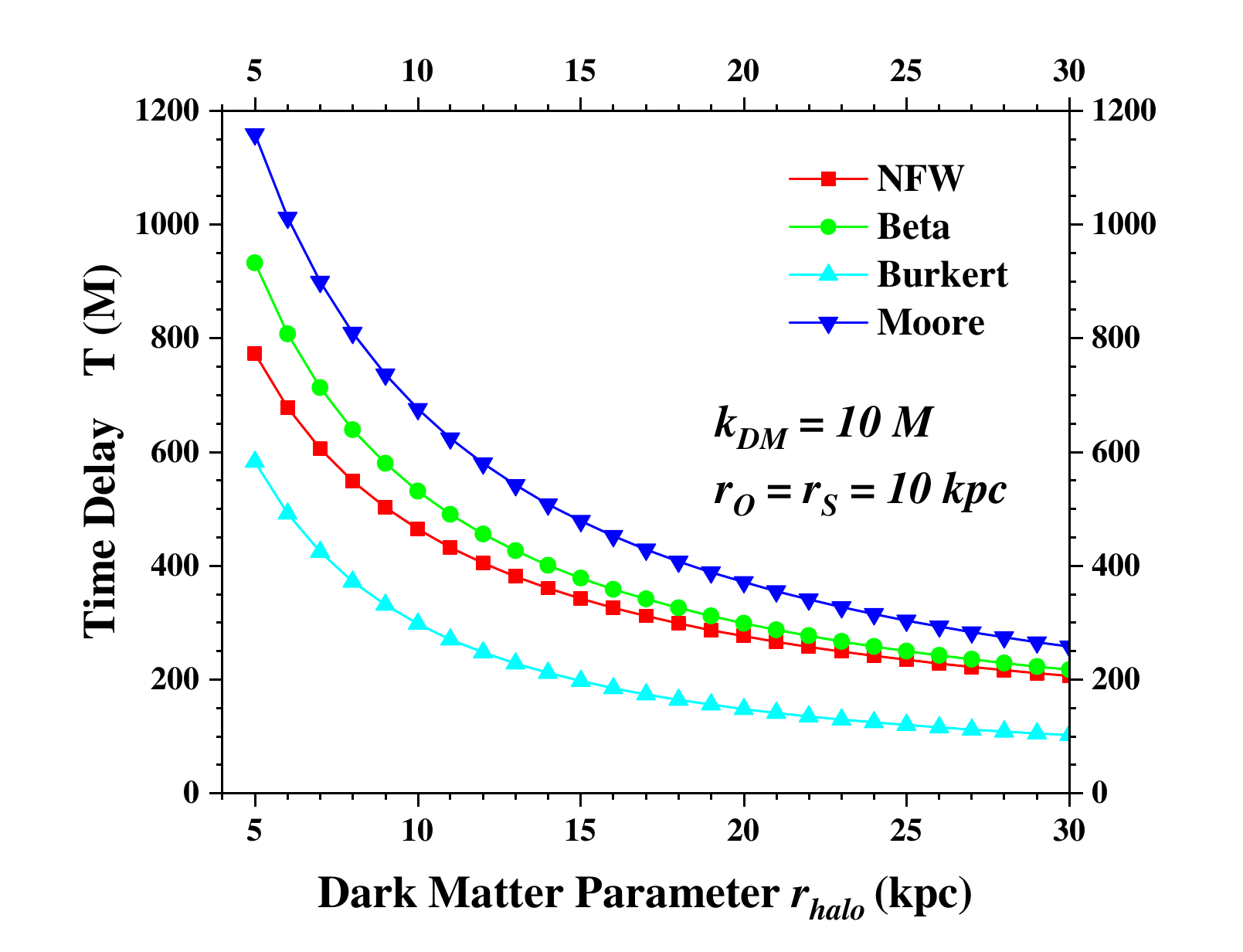}
	\caption{The time delay of light in the gravitational lensing of black hole surrounded by NFW, Beta, Burkert, Moore dark matter halos with dark matter parameters $k_{\text{DM}}$ and $r_{\text{halo}}$ varying. The left panel shows the variation of time delay when dark matter characteristic mass changes from $k_{\text{DM}} = 0$ to  $k_{\text{DM}} = 30M$ (with the dark matter halo characteristic scale chosen to be $r_{\text{halo}} = 10 \text{kpc}$). The right panel gives the tendency of time delay with dark matter halo characteristic scale increasing from $r_{\text{halo}} = 5 \text{kpc}$ to $r_{\text{halo}} = 30 \text{kpc}$ (where we have fixed the dark matter parameter $k_{\text{DM}} = 10M$). In this figure, the source and observer distance are selected to be $r_{\text{O}}=r_{\text{S}}=10 \ \text{kpc}$). }
	\label{figure3}
\end{figure*}

The numerical time delay results for black holes immersed in NFW, Beta, Burkert, Moore dark matter halos are presented in figure \ref{figure2} with several typical dark matter halo parameters. This figure shows the time delay of light for a static observer at $r_{\text{O}}=10^{10} M \sim 10 \text{kpc}$ with the varying of light source position $r_{\text{S}}$. The vertical axis represents the time delay results in unit of black hole mass $M$, and the horizontal axis denotes the light source radius $r_{\text{S}}$ measured in kpc. The characteristic mass of each dark matter halo models is selected as $k_{\text{DM}}=M$, $k_{\text{DM}}=5M$ $k_{\text{DM}}=10M$, and the dark matter halo scale is selected to be $r_{\text{halo}}=10^{10} M \sim 10 \ \text{kpc}$ and $r_{\text{halo}}=2\times 10^{10} M \sim 20 \ \text{kpc}$ respectively. Furthermore, to compare and emphasize the effects of dark matter halos, we illustrate the time delay of light for pure black hole cases without dark matter halos using black curves. Additionally, we also include analytical Shapiro time delay results $\Delta T = 2M\big(1+\ln(\frac{r_{\text{O}}r_{\text{S}}}{r_{0}^{2}})\big)$ in the weak field approximation for pure black hole cases 
\footnote{
The classical Shapiro time delay results in the weak field approximation, which is $\Delta T_{\text{classical Shapiro}} = 4M\big(1+\ln(\frac{r_{\text{A}}r_{\text{B}}}{r_{0}^{2}})\big)$, correspond to the light beams emitted from $A$ to $B$ and then get reflected from $B$ to $A$. However, in the gravitational lensing observations, the light beams are not get reflected, then the time delay of light should be the classical Shapiro time delay divided by a factor two, namely $\Delta T = \Delta T_{\text{classical Shapiro}} / 2 = 2M\big(1+\ln(\frac{r_{\text{O}}r_{\text{S}}}{r_{0}^{2}})\big)$.
\label{footnote4} }. 
From this figure, all the four aforementioned dark matter halos (the NFW, Beta, Burkert, Moore dark matter halos) have the potential to significantly enhance the time delay of light in the gravitational lensing of central supermassive black holes, and this enhancing effect becomes more pronounced as the dark matter halo characteristic mass $k_{\text{DM}}$ increases. Among the four dark matter halo models, the Moore model yields the largest time delay when the light source is not excessively far away from the observer, while the isothermal Beta model leads to the largest time delay for distant luminous source. The critical point occurs at approximately $r_{\text{O}} \sim 35-40 \ \text{kpc}$, both for parameters $r_{\text{halo}} = 10$ kpc and $r_{\text{halo}} = 20$ kpc. In all scenarios examined here, the time delay of light for pure black hole cases agree with the analytical Shapiro time delay results obtained within the weak field approximation.

In figure \ref{figure3}, we plot the time delay results for a static observer and light source with the variation of dark matter halo mass parameter $k_{\text{DM}}$ and dark matter halo scale parameter $r_{\text{halo}}$. This distance of observer and light source is chosen as $r_{\text{O}}=r_{\text{S}} = 10 \ \text{kpc}$. The left panel of figure \ref{figure3} shows that the time delay of light grows almost linearly as the dark matter halo characteristic mass $k_{\text{DM}}$ increases, for NFW, Beta, Burkert, Moore models. The natural explanation for this results is that, when the dark matter halo scale remains constant, a dark matter halo with a larger characteristic mass $k_{\text{DM}} = \rho_{0} \cdot r_{\text{halo}}^3$ could produce stronger gravitational field, therefore resulting in an increased time delay. Furthermore, the linear behavior of time delay results can also be explained from the analytical expansion of time delay. In appendix \ref{appendix1}, we provide the analytical expressions on time delay of light in the weak gravitational field limit for black holes surrounded by NFW and Beta dark matter halos, and the results show that the dark matter contributions to time delay are proportional to the dark matter characteristic mass $k_{\text{DM}}$ (see equations (\ref{time delay analytical NFW}) and (\ref{time delay analytical NFW}) in appendix \ref{appendix1}). Additionally, the right panel of figure \ref{figure3} illustrates that the time delay could reduce rapidly as dark matter halo scale $r_{\text{halo}}$ grows, for a constant dark matter halo mass. This is because, for a larger dark matter halo scale, the dark matter is more sparsely distributed and the gravitational field generated by dark matter halo is more weakly, resulting in an reduction of time delay.

In addition to the time delay, it is also interesting to explore other observables in gravitational lensing observations. In appendix \ref{appendix2}, we calculate the gravitational deflection angle of light for supermassive black holes surrounded by dark matter halos. The results of gravitational deflection angle exhibit a similar tendency as those for time delay presented in this section. A sparsely distributed dark matter distribution (with a small dark mater mass and a large halo scale) has tiny contribution to the gravitational deflection angle. Conversely, when the dark matter distribution becomes denser (with larger dark matter mass and smaller dark matter halo scale), its influences on gravitational deflection are significantly enhanced. Furthermore, the numerical results of gravitational deflection angle in Appendix \ref{appendix2} indicate that, under typical galactic dark matter parameters ($k_{\text{DM}}=10M$ and $r_{\text{halo}}=10$ kpc), the impacts of dark matter halos on the gravitational deflection angle seem to be less pronounced, compared with the influences on time delay of light. Particularly, in figure \ref{figure5}, under these typical galactic dark matter parameters ($k_{\text{DM}}=10M$ and $r_{\text{halo}}=10$ kpc), the time delay of light obtained using different dark matter halo models have notable differences, which may be potentially examined by astrophysical observations in galaxies (such as the gravitational lensing observations of Sgr*A in our Milky Way Galaxy). In contract, the gravitational deflection angles calculated from different dark matter halos models can't be effectively distinguished under these circumstances (see figure \ref{figure6} in Appendix \ref{appendix2} for detailed results).

\subsection{Time Delay Results in the Strong Gravitational Field Cases \label{sec:4b}}

\begin{figure*}
	\includegraphics[width=0.5\textwidth]{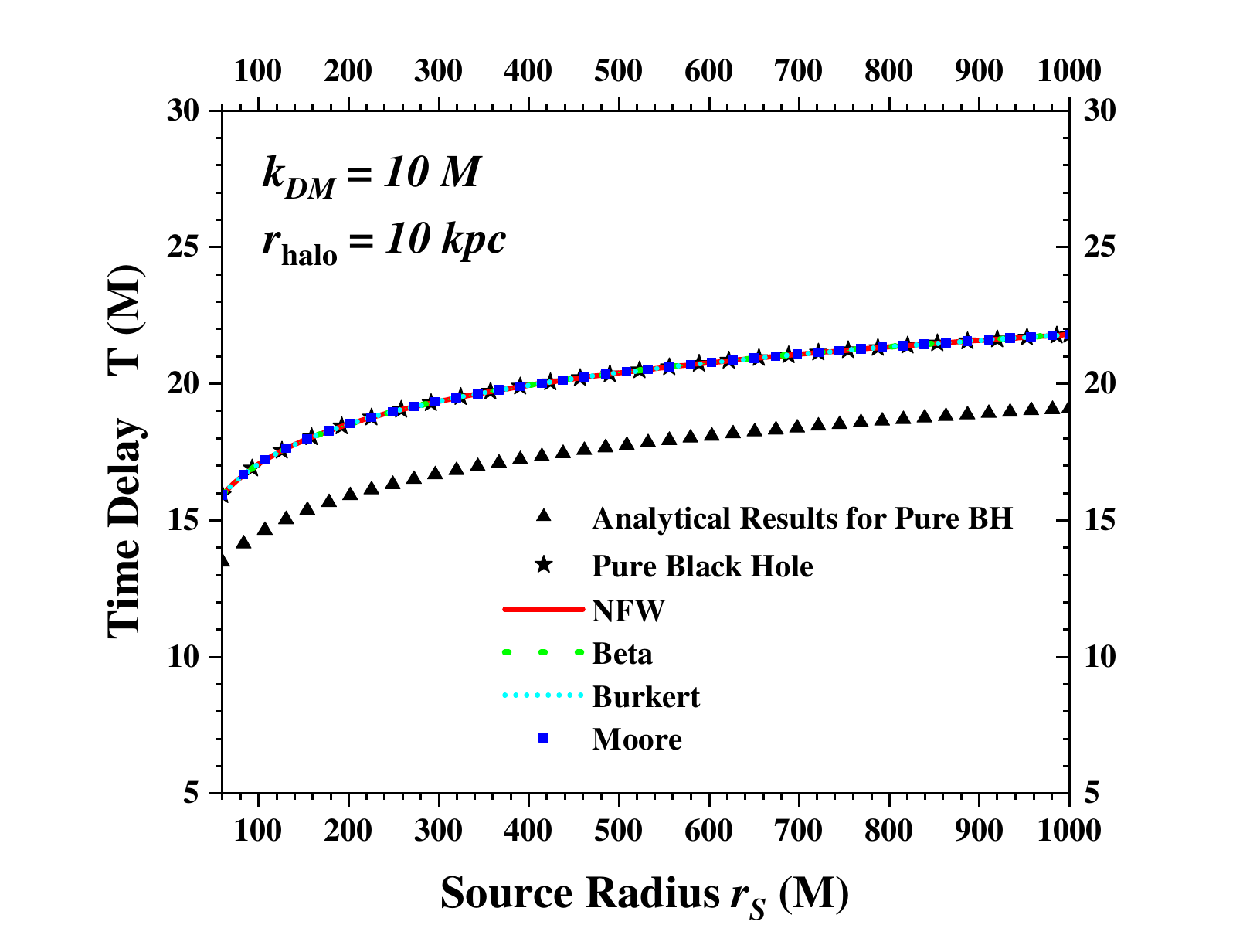}
	\includegraphics[width=0.5\textwidth]{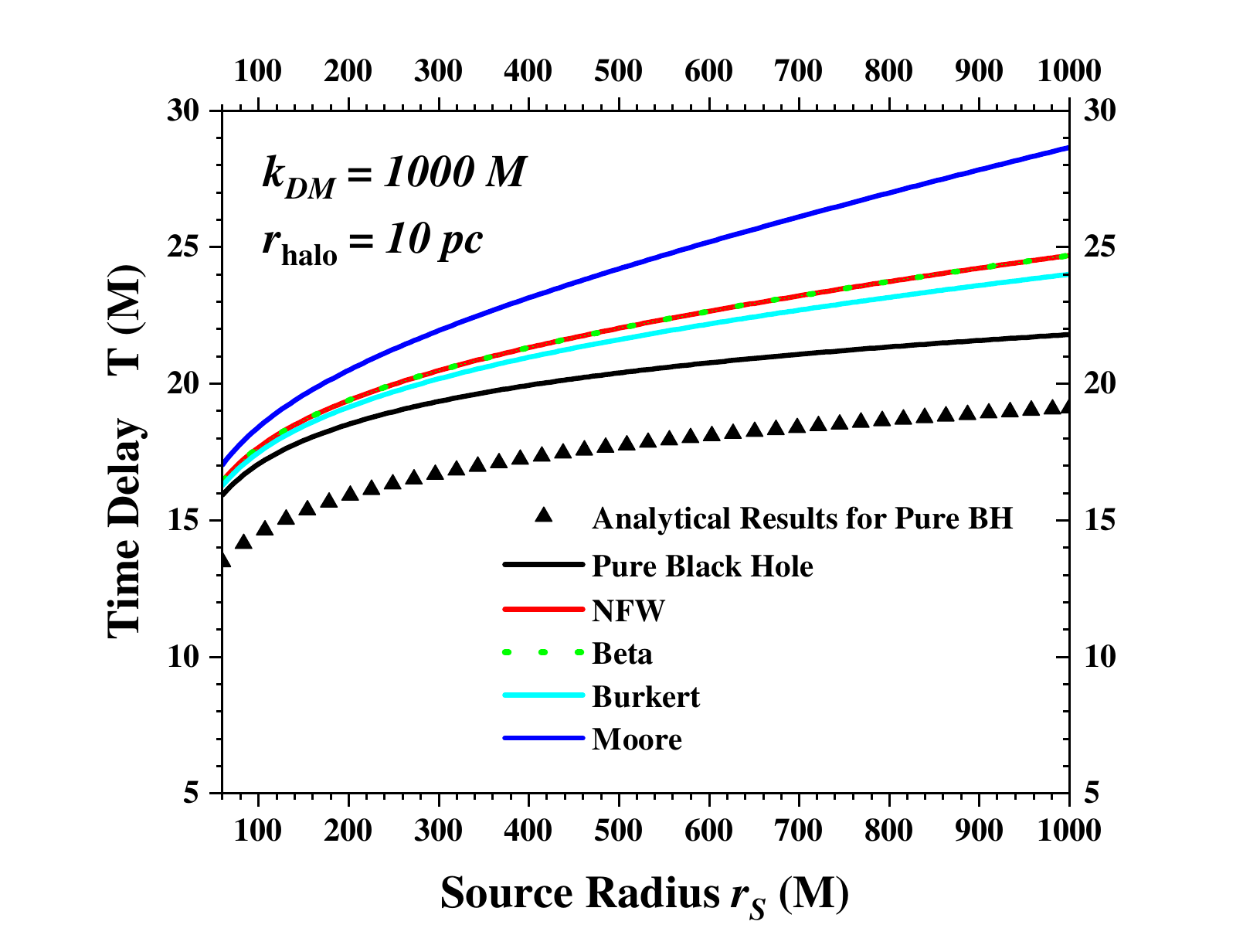}
	\caption{The time delay of light in gravitational lensing of black hole surrounded by dark matter halos in strong gravitational field cases. This figure presents numerical results calculated using the geodesic approach in equation (\ref{time delay for null geodesic}) (obtained by solving the trajectories of null geodesics). The numerical results presented here are obtained within the NFW, Bata, Burkert, Moore dark matter halos models. The dark matter characteristic mass is set as $k_{\text{DM}}=10M$ in the let panel and $k_{\text{DM}}=1000M$ in the right panel. The dark matter halo scale is selected as $r_{\text{halo}}=10^{10} M \sim 10$ kpc in the left panel and $r_{\text{halo}}= 10^{7} M \sim 10$ pc in the right panels, respectively. The time delay of light in gravitational lensing of pure black holes without the presence of dark matter halos is illustrated using black curves for comparisons. Additionally, the analytical results of Shapiro time delay $\Delta T = 2M\big(1+\ln(\frac{r_{\text{O}}r_{\text{S}}}{r_{0}^{2}})\big)$ for the pure black hole cases in the weak field approximation are also included. In this figure, the observer's location is at $r_{\text{O}}=100 M$, and the position of light source varies from $r_{\text{S}} = 60 M$ to $r_{\text{S}} = 1000 M$. The impact parameter of the photon orbit is fixed as $b=10 M$ (where the photon orbits are very close to the central black hole, subjecting to strong gravitational field / gravitational effects). }
	\label{figure4}
\end{figure*}

Theoretically, the exploration of dark matter effects on the time delay of light in strong gravitational cases is also of great significance. To present the time delay results in the strong gravitational cases for various dark matter halo models, it is necessary to position the light source and observer very close to the central black hole. Figure \ref{figure4} shows the numerical calculation of time delay in strong gravitational cases. In this figure, we choose the observer's position as $r_{\text{O}}=100 M$, and the position of light source varies from $r_{\text{S}} = 60 M$ to $r_{\text{S}} = 1000 M$. The left panel illustrates the time delay results for typical galactic dark matter parameters (which are closely connected with astrophysical gravitational lensing observations in galaxies), with the characteristic mass and halo scale for dark matter to be $k_{\text{DM}} = 10 M$ and $r_{\text{halo}}$ = 10 kpc. Unfortunately, under such parameters, there are no obvious differences in time delays calculated within NFW, Beta, Burkert, Mooore dark matter halos. The curves correspond to different dark matter halo models overlapped with each other in the left panel. It seems not possible to distinguish these halo models from the comparison of time delay results in strong gravitational cases. This is mainly due to the dispersal nature of dark matter distribution in galaxies. In NFW, Beta, Burkert, Moore dark matter halo models --- derived from astrophysical observations or simulations --- the distribution of dark matter spans across the entire galaxy, with the dark matter halo scale more than $10$ kpc. Consequently, the amount of dark matter in the vicinity of black hole only accounts for a very small proportion. Therefore, the dark matter does not have non-negligible effects on time delay of light in the near black hole region (where the strong gravitational field limit must be considered). In the strong gravitational field cases, in order to efficiently distinguish the time delays of light from different dark matter models, the dark matter mass should be increased to $k_{\text{DM}} \sim 1000 M$, meanwhile the scale of dark matter halo should be reduced to $r_{\text{halo}} \sim 10^7 M \sim 10$ pc (which is 3 orders of magnitude smaller than the astrophysical galactic scale, i.e. 10 kpc), as presented in the right panel of figure \ref{figure4}. It can be explained that a denser dark matter distribution (with a smaller scale and a larger mass) would result in stronger influences on time delays. Therefore, decreasing the size of the dark matter halo and increasing its mass could significantly amplify the differences of time delays in various dark matter halo models. Additionally, we also include analytical Shapiro time delay results $\Delta T = 2M\big(1+\ln(\frac{r_{\text{O}}r_{\text{S}}}{r_{0}^{2}})\big)$ for the pure black hole cases in figure \ref{figure4}. Notably, in the strong gravitational field cases, numerical results on time delay for pure black hole cases do not agree with the analytical Shapiro's time delay (obtained from the weak field approximation).

\section{Summary \label{sec:5}}

In this work, the time delay of light in the gravitational lensing of supermassive black holes in the galaxy center surrounded by dark matter halos is studied. To accurately describe the dark matter distributions in galaxies, we employ several astrophysical phenomenological dark matter halo models in the present work, namely the NFW, Beta, Burkert, and Moore halo models. Notably, we use effective spacetime metrics to duel with the gravitational field of supermassive black holes combined with dark matter halos in galaxies. This treatment enables us to investigate the time delay of light in a simpler way, without invoking any complicated numerical calculations and techniques in the astrophysical gravitational lensing. 

We calculate and compare the time delay of light for different dark matter halo models in both weak gravitational field cases (which are closely related to the gravitational lensing observations of black holes in galaxies) and strong gravitational field cases (for theoretical interests). The numerical results in the weak gravitational field cases demonstrate that, for a typical choice of characteristic length scales in the gravitational lensing of black holes in galaxies (with the black hole mass $M=10^{7} M_{\odot}$, dark matter holo scale $r_{\text{halo}} \sim $ 10 kpc, the observer radius to galaxy center $r_{\text{O}} \sim D_{\text{OL}} \sim 10$ kpc, the impact parameter in photon orbit $b \sim 0.1$ pc and Einstein ring $\theta_{E} \sim \ \text{arcsec}$), the NFW, Beta, Burkert, Moore dark matter halo models exhibit a significant enhancement on the time delay of light. This enhancing effect becomes more pronounced with an increasing dark matter halo characteristic mass $k_{\text{DM}}$. Furthermore, the time delay of light grows almost linearly as the dark matter characteristic mass increases, and this linear behavior can be explained from the analytical expansion of time delay in the appendix. Among these four dark matter halo models, the Moore model yields the largest time delay when the light source distance is not sufficiently far, while the isothermal Beta model leads to the largest time delay for distant luminous source. Additionally, the time delay can reduce rapidly as dark matter halo scale $r_{\text{halo}}$ grows (for a constant dark matter halo mass). A larger dark matter halo corresponds to a more sparse distribution of dark matter and a weaker gravitational field, therefore the reduction of time delay. In the strong gravitational field cases, for typical dark matter parameters in galaxies ($k_{\text{DM}} \sim 10M$ and $r_{\text{halo}} \sim $ 10 kpc), the dark matter halo effects on time delay of light are very weak, numerical results obtained within various dark matter halo models exhibit no obvious differences. To make the dark matter effects stronger and to efficiently distinguish the time delays of light from different dark matter models, the dark matter halo mass must be greatly increased and the dark matter halo scale should be dramatically reduced (at the order of $k_{\text{DM}} \sim 1000 M$ and $r_{\text{halo}} \sim 10^7 M \sim 10$ pc).

The treatment employed in the present study (with the effective spacetime metric to address the gravitational field of supermassive black holes surrounded by dark matter halos and the numerical scheme to obtain the time delay by solving the null geodesics) can be easily extended to a number of galactic and dark matter halo models. Our current work serves as a preliminary investigation on this field. In the future, the dark matter halo effects on time delay of light could be more deeply studied using alternative dark matter halo models and the incorporation of other characteristic length scales in galaxy and galaxy clusters.

\appendix

\section{Analytical Result of Time Delay in the Weak Gravitational Field Limit} \label{appendix1}

In this appendix, analytical results of the gravitational time delays for black holes in dark matter halos are presented. These analytical results enable us to give a explanation on the qualitative behaviors of time delay results with respect to the scale and characteristic mass of dark matter halos, especially those shown in figure \ref{figure3}, providing better understanding on the behavior of time delay of light in the presence dark matter halo. 

In the following, we present analytical expansions on time delay of light in the weak gravitational field limit for black holes surrounded by NFW and Beat dark matter halos. This can be achieved through two main procedures. Firstly, we apply asymptotic expansions to the effective spacetime metrics in equations (\ref{NFW}) and (\ref{Beta}) in the large distance limit. Subsequently, we evaluate the integral in time delay by further expanding the integration function in equation (\ref{time delay for null geodesic}) using some small parameters, similar to the derivation of Shapiro time delay result \cite{Weinberg1972}. However, for black holes surrounded by Burkert and Moore dark matter models, the asymptotic non-flatness nature of the effective metrics obstruct the asymptotic expansions in the large distance limit. Therefore, the analytical expression of time delay for black hole surrounded by Burkert and Moore dark matter models are not given in this section.

The asymptotic expansions of the effective metrics for black holes surrounded by NFW, Beta dark matter halos in the large distance limit $r\to\infty$ are given by
\begin{subequations}
\begin{eqnarray}
	f_{\text{NFW}}(r)  & = & (1+x)^{ -\frac{8 \pi k_{\text{DM}}}{r} } - \frac{2M}{r}
	= 1 - \frac{2M}{r} - \frac{8 \pi k_{\text{DM}} }{r} \cdot \ln \frac{r}{r_{\text{halo}}} + O\bigg( \frac{M^{2}}{r^2}, \frac{k_{\text{DM}}^{2}}{r^2}, \frac{Mk_{\text{DM}}}{r^2} \bigg) \label{NFW asymptotic expansion}
	\\
	f_{\text{Beta}}(r) & = & e^{-\frac{8 \pi k_{\text{DM}}}{r} \sinh^{-1} x} - \frac{2M}{r}
	= 1 - \frac{2M}{r} - \frac{8 \pi k_{\text{DM}} }{r} \cdot \ln \frac{2r}{r_{\text{halo}}}
	  + O\bigg( \frac{M^{2}}{r^2}, \frac{k_{\text{DM}}^{2}}{r^2}, \frac{Mk_{\text{DM}}}{r^2} \bigg) \label{Beta asymptotic expansion}
\end{eqnarray}
\end{subequations}
The effective spacetime metrics for black holes surrounded by NFW and Beta dark matter halos given in equations (\ref{NFW}) and (\ref{Beta}) are asymptotically flat, so the validity of such expansions in the large distance limit is verified. However, the effective metrics for black hole surrounded by Burkert and Moore dark matter halos in equations (\ref{Burkert}) and (\ref{Moore}) are not asymptotically flat. It brings about a cutoff of distance $r_{\text{max}}$ such that the valid region of the metric expressions in equations (\ref{Burkert})-(\ref{Moore}) is $r<r_{\text{max}}$, as well as the presence of an additional rescaled factor $f_{\text{rescale}} = 1/f(r_{\text{max}})$, as we have explained in the footnote \ref{footnote1}). This may prevent us obtaining a asymptotic expansion of the effective spacetime metric in the large distance limit $r\to\infty$. The analytical evaluation of the integral in the time delay of light with the original metric expressions (\ref{Burkert})-(\ref{Moore}) (rather than their asymptotic expansions in the large distance limit) is quite challenging, so the analytical expressions of time delay for black hole surrounded by Burkert and Moore dark matter halos are not given in this appendix.

To further obtain the analytical results of time delay $\Delta T$ into a power series, we should also assign some small parameters during the analytic evaluation of equation (\ref{time delay for null geodesic}). In this work, we choose the small parameters
\begin{equation}
	\epsilon = \frac{2M}{r_{0}} \ll 1 \ \ \ \ \ \ 
	\epsilon' = \frac{k_{\text{DM}}}{r_{0}} \ll 1
\end{equation}
with the $r_{0}$ to be the closet distance to central black holes in the photon orbit. In many astrophysical gravitational lensing observations, the light rays are not very close to the central black holes such that the gravitational lensing can be treated in the weak gravitational limit. In such cases, the closet distance $r_{0}$ is usually much larger than the black hole mass $M$ and the dark matter characteristic mass $k_{\text{DM}}$, making the $\epsilon$ and $\epsilon'$ to be small parameters which justify the analytical expansions. Additionally, we can slightly transform the from of integral in the expression of time delay in equation (\ref{time delay for null geodesic})
\begin{eqnarray}
	\Delta T = T - T_{0} 
	& = & \int_{r_{0}}^{r_{\text{S}}} \frac{dr}{f(r)\sqrt{1-\frac{b^{2} f(r)}{r^{2}}}}
	+ \int_{r_{0}}^{r_{\text{O}}} \frac{dr}{f(r)\sqrt{1-\frac{b^{2} f(r)}{r^{2}}}}
	- \sqrt{r_{\text{S}}^{2}-r_{0}^{2}} - \sqrt{r_{\text{O}}^{2}-r_{0}^{2}} \nonumber
	\\
	& = &  \int_{r_{0}}^{r_{\text{S}}} 
	\frac{dr}{f(r)\sqrt{1-\frac{r_{0}^{2}}{r^{2}}\cdot\frac{f(r)}{f(r_{0})}}}
	+ \int_{r_{0}}^{r_{\text{O}}} \frac{dr}{f(r)\sqrt{1-\frac{r_{0}^{2}}{r^{2}}\cdot\frac{f(r)}{f(r_{0})}}}
	- \sqrt{r_{\text{S}}^{2}-r_{0}^{2}} - \sqrt{r_{\text{O}}^{2}-r_{0}^{2}}
	\label{time delay appendix}
\end{eqnarray}
In the second line, the relationship between impact parameter and closet distance in equation (\ref{impactparameter-closetdistence}) has been used. 

Using these asymptotic expansions of metrics in expression (\ref{NFW asymptotic expansion})-(\ref{Beta asymptotic expansion}), the integration function in the time delay can be reduced to
\begin{subequations}
\begin{eqnarray}
	\frac{1}{\sqrt{1-\frac{r_{0}^{2}}{r^{2}}\cdot \frac{f_{\text{NFW}}(r)}{f_{\text{NFW}}(r_{0})}}} 
	& = & \frac{1}{\sqrt{1-\frac{r_{0}^{2}}{r^{2}}\cdot\frac{1-\frac{2M}{r}-\frac{8\pi k_{\text{DM}}}{r}\cdot\ln\frac{r}{r_{\text{halo}}}+O(\epsilon^{2},\epsilon'^{2},\epsilon\epsilon')}{1-\frac{2M}{r_{0}}-\frac{8\pi k_{\text{DM}}}{r}\cdot\ln\frac{r_{0}}{r_{\text{halo}}}+O(\epsilon^{2},\epsilon'^{2},\epsilon\epsilon')}}} \nonumber
	\\
	& = & \frac{1}{\sqrt{1-\frac{r_{0}^{2}}{r^{2}} \cdot \big[ 1-\frac{2M}{r}-\frac{8\pi k_{\text{DM}}}{r}\cdot\ln\frac{r_{0}}{r_{\text{halo}}}-\frac{8\pi k_{\text{DM}}}{r}\cdot\ln\frac{r}{r_{0}} \big] \cdot \big[1+\frac{2M}{r_{0}}+\frac{8\pi k_{\text{DM}}}{r}\cdot\ln\frac{r_{0}}{r_{\text{halo}}}\big]+O(\epsilon^{2},\epsilon'^{2},\epsilon\epsilon')}} \nonumber
	\\
	& = & \frac{1}{\sqrt{1-\frac{r_{0}^{2}}{r^{2}} \cdot \big[ 1-\frac{2M}{r+r_{0}}\frac{r_{0}}{r}-\frac{8\pi k_{\text{DM}}}{r+r_{0}} \frac{r_{0}}{r}\cdot\ln\frac{r_{0}}{r_{\text{halo}}} +\frac{8\pi k_{\text{DM}}}{r+r_{0}}\frac{r_{0}^{2}}{r(r-r_{0})}\cdot\ln\frac{r}{r_{0}} \big] +O(\epsilon^{2},\epsilon'^{2},\epsilon\epsilon')}} \nonumber
	\\
	& = & \frac{1+\frac{M}{r+r_{0}}\frac{r_{0}}{r}+\frac{4\pi k_{\text{DM}}}{r+r_{0}} \frac{r_{0}}{r}\cdot\ln\frac{r_{0}}{r_{\text{halo}}} - \frac{4\pi k_{\text{DM}}}{r+r_{0}}\frac{r_{0}^{2}}{r(r-r_{0})}\cdot\ln\frac{r}{r_{0}} +O(\epsilon^{2},\epsilon'^{2},\epsilon\epsilon')}{\sqrt{1-\frac{r_{0}^{2}}{r^{2}}}}
	\\
	\frac{1}{\sqrt{1-\frac{r_{0}^{2}}{r^{2}}\cdot \frac{f_{\text{Beta}}(r)}{f_{\text{Beta}}(r_{0})}}}
	& = & \frac{1}{\sqrt{1-\frac{r_{0}^{2}}{r^{2}}\cdot\frac{1-\frac{2M}{r}-\frac{8\pi k_{\text{DM}}}{r}\cdot\ln\frac{2r}{r_{\text{halo}}}+O(\epsilon^{2},\epsilon'^{2},\epsilon\epsilon')}{1-\frac{2M}{r_{0}}-\frac{8\pi k_{\text{DM}}}{r}\cdot\ln\frac{2r_{0}}{r_{\text{halo}}}+O(\epsilon^{2},\epsilon'^{2},\epsilon\epsilon')}}} \nonumber
	\\
	& = & \frac{1}{\sqrt{1-\frac{r_{0}^{2}}{r^{2}} \cdot \big[ 1-\frac{2M}{r}-\frac{8\pi k_{\text{DM}}}{r}\cdot\ln\frac{2r_{0}}{r_{\text{halo}}}-\frac{8\pi k_{\text{DM}}}{r}\cdot\ln\frac{r}{r_{0}} \big] \cdot \big[1+\frac{2M}{r_{0}}+\frac{8\pi k_{\text{DM}}}{r}\cdot\ln\frac{2r_{0}}{r_{\text{halo}}}\big] +O(\epsilon^{2},\epsilon'^{2},\epsilon\epsilon')}} \nonumber
	\\
	& = & \frac{1}{\sqrt{1-\frac{r_{0}^{2}}{r^{2}} \cdot \big[ 1-\frac{2M}{r+r_{0}}\frac{r_{0}}{r}-\frac{8\pi k_{\text{DM}}}{r+r_{0}} \frac{r_{0}}{r}\cdot\ln\frac{2r_{0}}{r_{\text{halo}}} +\frac{8\pi k_{\text{DM}}}{r+r_{0}}\frac{r_{0}^{2}}{r(r-r_{0})}\cdot\ln\frac{r}{r_{0}} \big] +O(\epsilon^{2},\epsilon'^{2},\epsilon\epsilon')}} \nonumber
	\\
	& = & \frac{1+\frac{M}{r+r_{0}}\frac{r_{0}}{r}+\frac{4\pi k_{\text{DM}}}{r+r_{0}} \frac{r_{0}}{r}\cdot\ln\frac{2r_{0}}{r_{\text{halo}}} - \frac{4\pi k_{\text{DM}}}{r+r_{0}}\frac{r_{0}^{2}}{r(r-r_{0})}\cdot\ln\frac{r}{r_{0}} +O(\epsilon^{2},\epsilon'^{2},\epsilon\epsilon')}{\sqrt{1-\frac{r_{0}^{2}}{r^{2}}}}
\end{eqnarray}
\end{subequations}
where we have utilize the small parameters $\frac{2M}{r} \le \frac{2M}{r_{0}} = \epsilon \ll 1$, $\frac{k_{\text{DM}}}{r} \le \frac{k_{\text{DM}}}{r_{0}} = \epsilon' \ll 1$ in the derivation. With the above expansions over small parameters $\epsilon$, $\epsilon'$ for NFW and Beta dark matter models, the main integral in the time delay of light given in equation (\ref{time delay appendix}) can be evaluated
\footnote{The derivation of these expressions include evaluation of the definite integral $\int_{r_{0}}^{R} \frac{8\pi k_{\text{DM}}}{r} \ln\frac{r}{r_{0}} \frac{dr}{\sqrt{1-\frac{r_{0}^{2}}{r^{2}}}}$, which is not easy to get a simple the analytical result from basic calculus. However, we find an expression from the Wolfram Mathematica $\int_{r_{0}}^{R} \frac{1}{r} \ln\frac{r}{r_{0}} \frac{dr}{\sqrt{1-\frac{r_{0}^{2}}{r^{2}}}} \approx \frac{\pi^{2}}{24} -\frac{(\ln2)^{2}}{2}  + \frac{1}{2} \bigg(\ln \frac{R}{r_{0}} \bigg)^{2} + \ln \frac{R}{r_{0}} \cdot \ln \frac{1+\sqrt{1-\frac{r_{0}^{2}}{R^{2}}}}{2} $.} 
\begin{subequations}
\begin{eqnarray}
	\int_{r_{0}}^{R} \frac{dr}{f_{\text{NFW}}(r)\sqrt{1-\frac{r_{0}^{2}}{r^{2}}\cdot \frac{f_{\text{NFW}}(r)}{f_{\text{NFW}}(r_{0})}}} 
	& = & \int_{r_{0}}^{R} \bigg\{ \frac{1+\frac{M}{r+r_{0}}\frac{r_{0}}{r}+\frac{4\pi k_{\text{DM}}}{r+r_{0}} \frac{r_{0}}{r}\cdot\ln\frac{r_{0}}{r_{\text{halo}}} - \frac{4\pi k_{\text{DM}}}{r+r_{0}}\frac{r_{0}^{2}}{r(r-r_{0})}\cdot\ln\frac{r}{r_{0}} + O(\epsilon^{2},\epsilon'^{2},\epsilon\epsilon') }{\sqrt{1-\frac{r_{0}^{2}}{r^{2}}}} \nonumber
	\\
	&  & \ \ \ \ \ \ \ \cdot
	\bigg[ 1 + \frac{2M}{r} + \frac{8\pi k_{\text{DM}}}{r}\cdot\ln\frac{r_{0}}{r_{\text{halo}}} + \frac{8\pi k_{\text{DM}}}{r} \cdot \ln\frac{r}{r_{0}} + O(\epsilon^{2},\epsilon'^{2},\epsilon\epsilon') \bigg] \bigg\} dr \nonumber
	\\
	& \approx & \sqrt{R^{2}-r_{0}^{2}} + M \sqrt{\frac{R-r_{0}}{R+r_{0}}} + 2M\ln\bigg(\frac{R+\sqrt{R^{2}-r_{0}^{2}}}{r_{0}}\bigg)  \nonumber
	\\
	&   & + 4\pi k_{\text{DM}} \cdot \ln\frac{r_{0}}{r_{\text{halo}}} \cdot \sqrt{\frac{R-r_{0}}{R+r_{0}}} + 8\pi k_{\text{DM}} \cdot \ln\frac{r_{0}}{r_{\text{halo}}} \cdot \ln\frac{R+\sqrt{R^{2}-r_{0}^{2}}}{r_{0}}  \nonumber
	\\
	&   & + 4\pi k_{\text{DM}} \cdot \frac{R}{\sqrt{R^{2}-r_{0}^{2}}} \cdot \ln\frac{R}{r_{0}} - 4\pi k_{\text{DM}} \ln\frac{R+\sqrt{R^{2}-r_{0}^{2}}}{r_{0}} \nonumber
	\\
	&   & + 8\pi k_{\text{DM}} \bigg[ \frac{\pi^{2}}{24} -\frac{(\ln2)^{2}}{2}  + \frac{1}{2} \bigg(\ln \frac{R}{r_{0}} \bigg)^{2} + \ln \frac{R}{r_{0}} \cdot \ln \frac{1+\sqrt{1-\frac{r_{0}^{2}}{R^{2}}}}{2}  \bigg]
	\label{analytical evaluation NFW}
	\\
	\int \frac{dr}{f_{\text{Beta}}(r)\sqrt{1-\frac{r_{0}^{2}}{r^{2}}\cdot \frac{f_{\text{Beta}}(r)}{f_{\text{Beta}}(r_{0})}}}
	& = & \int \bigg\{ \frac{1+\frac{M}{r+r_{0}}\frac{r_{0}}{r}+\frac{4\pi k_{\text{DM}}}{r+r_{0}} \frac{r_{0}}{r}\cdot\ln\frac{2r_{0}}{r_{\text{halo}}} - \frac{4\pi k_{\text{DM}}}{r+r_{0}}\frac{r_{0}^{2}}{r(r-r_{0})}\cdot\ln\frac{r}{r_{0}} + O(\epsilon^{2},\epsilon'^{2},\epsilon\epsilon') }{\sqrt{1-\frac{r_{0}^{2}}{r^{2}}}} \nonumber
	\\
	&  & \ \ \ \ \ \cdot
	\bigg[ 1 + \frac{2M}{r} + \frac{8\pi k_{\text{DM}}}{r}\cdot\ln\frac{2r_{0}}{r_{\text{halo}}} + \frac{8\pi k_{\text{DM}}}{r} \cdot \ln\frac{r}{r_{0}} + O(\epsilon^{2},\epsilon'^{2},\epsilon\epsilon') \bigg] \bigg\} dr \nonumber
	\\
	& \approx & \sqrt{R^{2}-r_{0}^{2}} + M \sqrt{\frac{R-r_{0}}{R+r_{0}}} + 2M \cdot \ln\frac{R+\sqrt{R^{2}-r_{0}^{2}}}{r_{0}}  \nonumber
	\\
	&   & + 4\pi k_{\text{DM}} \cdot \ln\frac{2r_{0}}{r_{\text{halo}}} \cdot \sqrt{\frac{R-r_{0}}{R+r_{0}}} + 8\pi k_{\text{DM}} \cdot \ln\frac{2r_{0}}{r_{\text{halo}}} \cdot \ln\frac{R+\sqrt{R^{2}-r_{0}^{2}}}{r_{0}}  \nonumber
	\\
	&   & + 4\pi k_{\text{DM}} \cdot \frac{R}{\sqrt{R^{2}-r_{0}^{2}}} \cdot \ln\frac{R}{r_{0}} - 4\pi k_{\text{DM}} \cdot \ln\frac{R+\sqrt{R^{2}-r_{0}^{2}}}{r_{0}} \nonumber
	\\
	&   & + 8\pi k_{\text{DM}} \bigg[ \frac{\pi^{2}}{24} -\frac{(\ln2)^{2}}{2}  + \frac{1}{2} \bigg(\ln \frac{R}{r_{0}} \bigg)^{2} + \ln \frac{R}{r_{0}} \cdot \ln \frac{1+\sqrt{1-\frac{r_{0}^{2}}{R^{2}}}}{2}  \bigg] 
	\label{analytical evaluation Beta}
\end{eqnarray}
\end{subequations}
Eventually, the time delay of light for gravitational lensing of black hole surrounded by NFW and Beta dark matter halo models in the weak gravitational field cases can be obtained using equations (\ref{time delay appendix}) and (\ref{analytical evaluation NFW})-(\ref{analytical evaluation Beta})
\begin{subequations}
\begin{eqnarray}
	\Delta T_{\text{NFW}} 
	& \approx & M \bigg( \sqrt{\frac{r_{\text{S}}-r_{0}}{r_{\text{S}}+r_{0}}} + \sqrt{\frac{r_{\text{O}}-r_{0}}{r_{\text{O}}+r_{0}}} \bigg) 
	+ 2M \cdot \ln \frac{(r_{\text{S}}+\sqrt{r_{\text{S}}^{2}-r_{0}^{2}})(r_{\text{O}}+\sqrt{r_{\text{O}}^{2}-r_{0}^{2}})}{r_{0}^{2}}  \nonumber
	\\
	&   & + 4\pi k_{\text{DM}} \cdot \bigg( \ln\frac{r_{0}}{r_{\text{halo}}} \bigg) \cdot \bigg( \sqrt{\frac{r_{\text{S}}-r_{0}}{r_{\text{S}}+r_{0}}} + \sqrt{\frac{r_{\text{O}}-r_{0}}{r_{\text{O}}+r_{0}}} \bigg)
	+ 8\pi k_{\text{DM}} \cdot \ln\frac{r_{0}}{r_{\text{halo}}} \cdot \ln \frac{(r_{\text{S}}+\sqrt{r_{\text{S}}^{2}-r_{0}^{2}})(r_{\text{O}}+\sqrt{r_{\text{O}}^{2}-r_{0}^{2}})}{r_{0}^{2}}  \nonumber
	\\
	&   & + 4\pi k_{\text{DM}} \bigg( \frac{r_{\text{S}}}{\sqrt{r_{\text{S}}^{2}-r_{0}^{2}}} \cdot \ln\frac{r_{\text{S}}}{r_{0}} + \frac{r_{\text{O}}}{\sqrt{r_{\text{O}}^{2}-r_{0}^{2}}} \cdot \ln\frac{r_{\text{O}}}{r_{0}} \bigg) 
	- 4\pi k_{\text{DM}} \cdot \ln \frac{(r_{\text{S}}+\sqrt{r_{\text{S}}^{2}-r_{0}^{2}})(r_{\text{O}}+\sqrt{r_{\text{O}}^{2}-r_{0}^{2}})}{r_{0}^{2}} \nonumber
	\\
	&   & + 8\pi k_{\text{DM}} \bigg[ \frac{\pi^{2}}{12} -(\ln2)^{2}  + \frac{1}{2} \bigg(\ln \frac{r_{\text{S}}}{r_{0}} \bigg)^{2} + \frac{1}{2} \bigg(\ln \frac{r_{\text{O}}}{r_{0}} \bigg)^{2} + \ln \frac{r_{\text{S}}}{r_{0}} \cdot \ln \frac{1+\sqrt{1-\frac{r_{0}^{2}}{r_{\text{S}}^{2}}}}{2} + \ln \frac{r_{\text{O}}}{r_{0}} \cdot \ln \frac{1+\sqrt{1-\frac{r_{0}^{2}}{r_{\text{O}}^{2}}}}{2}  \bigg] \nonumber
	\\
	\label{time delay analytical NFW}
	\\
	\Delta T_{\text{Beta}} 
	& \approx & M \bigg( \sqrt{\frac{r_{\text{S}}-r_{0}}{r_{\text{S}}+r_{0}}} + \sqrt{\frac{r_{\text{O}}-r_{0}}{r_{\text{O}}+r_{0}}} \bigg) 
	+ 2M \cdot \ln \frac{(r_{\text{S}}+\sqrt{r_{\text{S}}^{2}-r_{0}^{2}})(r_{\text{O}}+\sqrt{r_{\text{O}}^{2}-r_{0}^{2}})}{r_{0}^{2}}  \nonumber
	\\
	&   & + 4\pi k_{\text{DM}} \cdot \bigg( \ln\frac{2r_{0}}{r_{\text{halo}}} \bigg) \cdot \bigg( \sqrt{\frac{r_{\text{S}}-r_{0}}{r_{\text{S}}+r_{0}}} + \sqrt{\frac{r_{\text{O}}-r_{0}}{r_{\text{O}}+r_{0}}} \bigg)
	+ 8\pi k_{\text{DM}} \cdot \ln\frac{2r_{0}}{r_{\text{halo}}} \cdot \ln \frac{(r_{\text{S}}+\sqrt{r_{\text{S}}^{2}-r_{0}^{2}})(r_{\text{O}}+\sqrt{r_{\text{O}}^{2}-r_{0}^{2}})}{r_{0}^{2}}  \nonumber
	\\
	&   & + 4\pi k_{\text{DM}} \bigg( \frac{r_{\text{S}}}{\sqrt{r_{\text{S}}^{2}-r_{0}^{2}}} \cdot \ln\frac{r_{\text{S}}}{r_{0}} + \frac{r_{\text{O}}}{\sqrt{r_{\text{O}}^{2}-r_{0}^{2}}} \cdot \ln\frac{r_{\text{O}}}{r_{0}} \bigg) 
	- 4\pi k_{\text{DM}} \cdot \ln \frac{(r_{\text{S}}+\sqrt{r_{\text{S}}^{2}-r_{0}^{2}})(r_{\text{O}}+\sqrt{r_{\text{O}}^{2}-r_{0}^{2}})}{r_{0}^{2}} \nonumber
	\\
	&   & + 8\pi k_{\text{DM}} \bigg[ \frac{\pi^{2}}{12} -(\ln2)^{2}  + \frac{1}{2} \bigg(\ln \frac{r_{\text{S}}}{r_{0}} \bigg)^{2} + \frac{1}{2} \bigg(\ln \frac{r_{\text{O}}}{r_{0}} \bigg)^{2} + \ln \frac{r_{\text{S}}}{r_{0}} \cdot \ln \frac{1+\sqrt{1-\frac{r_{0}^{2}}{r_{\text{S}}^{2}}}}{2} + \ln \frac{r_{\text{O}}}{r_{0}} \cdot \ln \frac{1+\sqrt{1-\frac{r_{0}^{2}}{r_{\text{O}}^{2}}}}{2}  \bigg] \nonumber
	\\
	\label{time delay analytical Beta}
\end{eqnarray}
\end{subequations}
with $r_{\text{S}}$ and $r_{\text{O}}$ to be the radial coordinate of light source and observer. The two terms in the first line of equations (\ref{time delay analytical NFW})-(\ref{time delay analytical Beta}) reflect the central black hole's gravitational effects on time delay, while other terms reflect to the dark matter halo contributions to time delay of light. From these expressions, it is clearly to observe that the dark matter's contributions are proportional to the dark matter characteristic mass $k_{\text{DM}}$, which can explain the linear behavior of time delay results given in the left panel of figure \ref{figure3}. From the two terms in second line of equations (\ref{time delay analytical NFW})-(\ref{time delay analytical Beta}), the dark matter halo scale $r_{\text{halo}}$ contribute to the time delay of light via $\ln(r_{0}/r_{\text{halo}})$ (for NFW model) or $\ln(2r_{0}/r_{\text{halo}})$ (for Beta model). For a fixed dark matter mass, when the dark matter halo scare increases, their contributions on time delay reduced as a factor of $\ln(1/r_{\text{halo}})$, which explain the decreasing trend of numerical results presented in the right panel of figure \ref{figure3}. Furthermore, in the absence of dark matter, the results recover to the famous Shapiro time delay result $\Delta T = \Delta_{\text{classical Shapiro}}/2 = 2M \big( 1+ \ln \frac{r_{\text{S}}r_{\text{O}}}{r_{0}^{2}} \big)$ for the distant light source and observer ($r_{\text{S}} \gg r_{0}$, $r_{\text{O}} \gg r_{0}$).

\section{Numerical Results on Gravitational Deflection Angle of Light} \label{appendix2}

In this appendix, the gravitational deflection angle of light in the gravitational lensing of supermassive black holes surrounded by dark matter halos are presented. The gravitational deflection angles of light can be calculated in a similar method for time delay, which has been given in section \ref{sec:3}. We calculated the gravitational deflection angles for both weak gravitational field cases (with typical dark matter parameters in galactic gravitational lensing) and strong gravitational field cases.

\begin{figure*}
	\centering
	\includegraphics[width=0.75\textwidth]{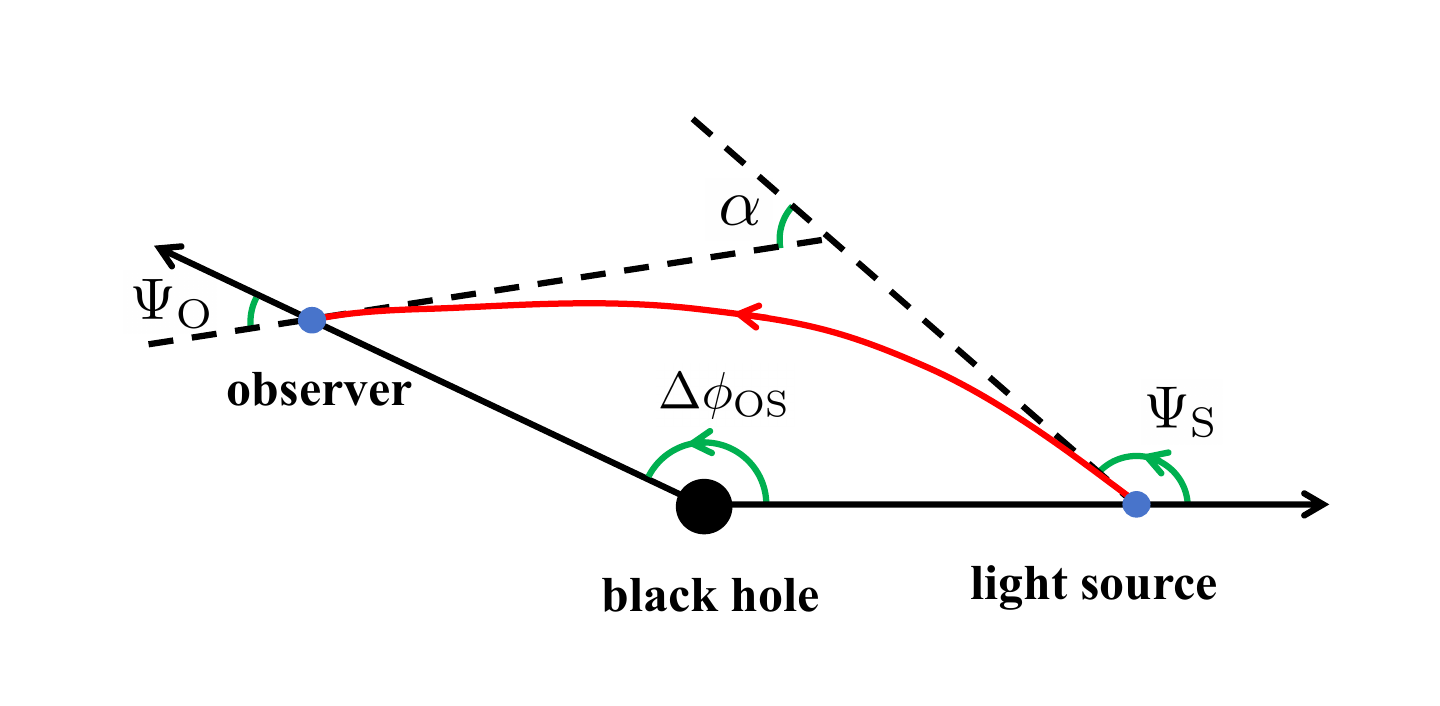}
	\caption{The gravitational deflection angle of light for finite distance light source and observer. In the thin lens approximation, the gravitational deflection angle $\alpha \equiv \Psi_{\text{O}} - \Psi_{\text{S}} - \Delta \phi_{\text{OS}}$ can be understand in a simple and straightforward way. For the quadrangle depicted in this figure, the exterior angles at each point are $\Psi_{\text{S}}$, $\alpha$, $\pi-\Psi_{\text{O}}$, $\pi - \Delta \phi_{\text{OS}}$, respectively. The thin lens approximation assumes that the spacetime is flat except for the location of thin lens (the central supermassive black hole), so the sum of exterior angles equals $2\pi$, which justify the definition of gravitational deflection angle $\alpha \equiv \Psi_{\text{O}} - \Psi_{\text{S}} - \Delta \phi_{\text{OS}}$. However, it should be noted that this definition given by Ishihara \emph{et al.} is generally valid in gravitational lensing beyond the thin lens approximation, see references \cite{Ishihara2016,Takizawa2020b} for more discussions on this issue}
	\label{figure5}
\end{figure*}

In the gravitational lensing, when the light source and observer are located at a finite distance region to the central black hole (with the radial coordinate $r=r_{\text{S}}$ and $r=r_{\text{O}}$, respectively), the gravitational deflection angle of light is defined through \cite{Ishihara2016}
\begin{equation}
	\alpha \equiv \Psi_{\text{O}} - \Psi_{\text{S}} - \Delta \phi_{\text{OS}}
	\label{gravitational deflection angle}
\end{equation}
The notation $\Psi_{\text{O}}$, $\Psi_{\text{S}}$ denote the angles between light ray propagation direction and the radial direction (that are measured at the observer's position and source position), and $\Delta \phi_{\text{OS}}$ is the variation of azimuthal angle $\phi$ in the light trajectories. Particularly, for a spherically symmetric spacetime described by equation (\ref{spacetime metric spherically symmetric}), the angle between light ray propagation direction and the radial direction is uniquely determined by spacetime metric \cite{Ishihara2016}, via
\begin{equation}
	\sin \Psi = \frac{b \cdot \sqrt{f(r)}}{r} \label{angle psi}
\end{equation} 
In the thin lens approximation, this definition of gravitational deflection angle for the finite distance source and observer is illustrated in figure \ref{figure5}.

In a spherically symmetric spacetime, the variation of azimuthal angle $\phi$ in the light trajectory is derived from the reduced differential equation (\ref{reduced differential equation}) for geodesics 
\begin{equation}
	\frac{dr}{d\phi} = \frac{dr}{d\lambda} \cdot \frac{d\lambda}{d\phi}
	= \pm r^{2} \sqrt{\frac{1}{b^{2}}-\frac{f(r)}{r^{2}}}
\end{equation}
where we have used the affine parameter $\epsilon=0$ for massless photon, the conserved energy parameter $E = f(r) \frac{dt}{d\lambda}$, conserved angular momentum $J = r^2 \sin^{2}\theta \frac{d\phi}{d\lambda}$, and the definition of impact parameter $b\equiv |J/E|$. The plus and minus sign $\pm$ can be determined similar to the cases of time delay
\begin{subequations}
	\begin{eqnarray}
		&& \frac{dr}{d\phi} = - r^{2} \sqrt{\frac{1}{b^{2}}-\frac{f(r)}{r^{2}}} < 0 \ \ \ \text{photon moving from source position $r=r_{\text{S}}$ to the tuning point $r=r_{0}$} 
		\nonumber
		\\
		&& \frac{dr}{d\phi} = r^{2} \sqrt{\frac{1}{b^{2}}-\frac{f(r)}{r^{2}}} > 0 \ \ \ \ \ \text{photon moving from the tuning point $r=r_{0}$ to observer position $r=r_{\text{O}}$}  
		\nonumber
	\end{eqnarray}
\end{subequations} 
With the expression of angle $\Psi$ in equation (\ref{angle psi}) and the variation of azimuthal angle, the gravitational deflection angle of light can be expressed as
\begin{eqnarray}
	\alpha = \Psi_{\text{O}} - \Psi_{\text{S}} - \Delta \phi_{\text{OS}}
	& = & \Psi_{\text{O}} - \Psi_{\text{S}} 
	+ \int_{r_{\text{S}}}^{r_{0}} \frac{d\phi}{dr} dr 
	+ \int_{r_{0}}^{r_{\text{O}}} \frac{d\phi}{dr} dr \nonumber
	\\
	& = & \arcsin \bigg( \frac{b \sqrt{f(r_{\text{R}})}}{r_{\text{R}}} \bigg)
	+ \arcsin \bigg( \frac{b \sqrt{f(r_{\text{S}})}}{r_{\text{S}}} \bigg) - \pi
	+ \int_{r_{0}}^{r_{\text{S}}} \frac{dr}{r^{2}\sqrt{\frac{1}{b^{2}}-\frac{f(r)}{r^{2}}}}
	+ \int_{r_{0}}^{r_{\text{O}}} \frac{dr}{r^{2}\sqrt{\frac{1}{b^{2}}-\frac{f(r)}{r^{2}}}} \nonumber
	\\
	\label{gravitational deflection angle for null geodesic}
\end{eqnarray}
As shown in figure \ref{figure5}, the angle $\Psi$ at the position of light source is an obtuse angle such that $\Psi_{\text{S}} = \pi - \arcsin \big( \frac{b \sqrt{f(r_{\text{S}})}}{r_{\text{S}}} \big)$.

\begin{figure*}
	\includegraphics[width=0.5\textwidth]{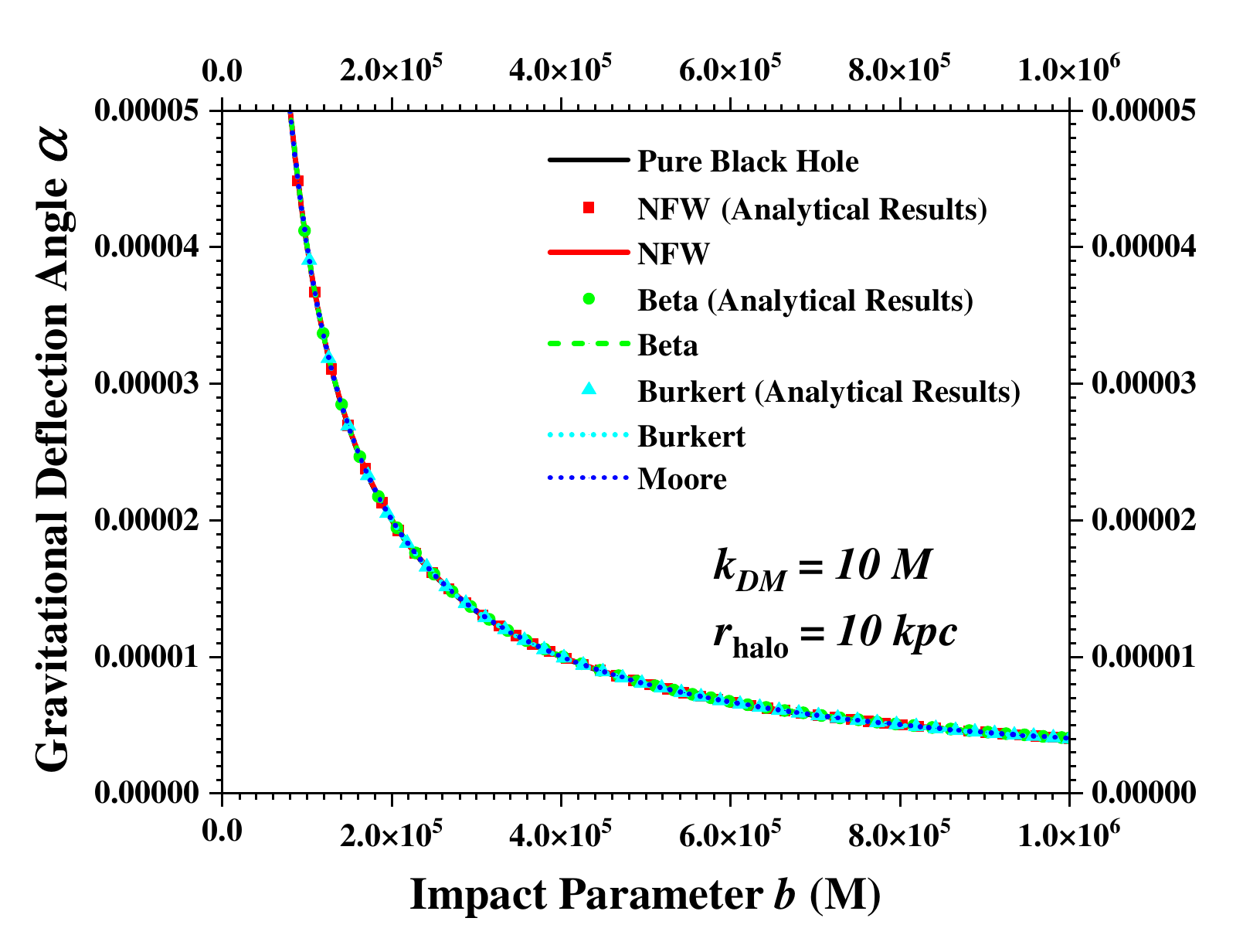}
	\includegraphics[width=0.5\textwidth]{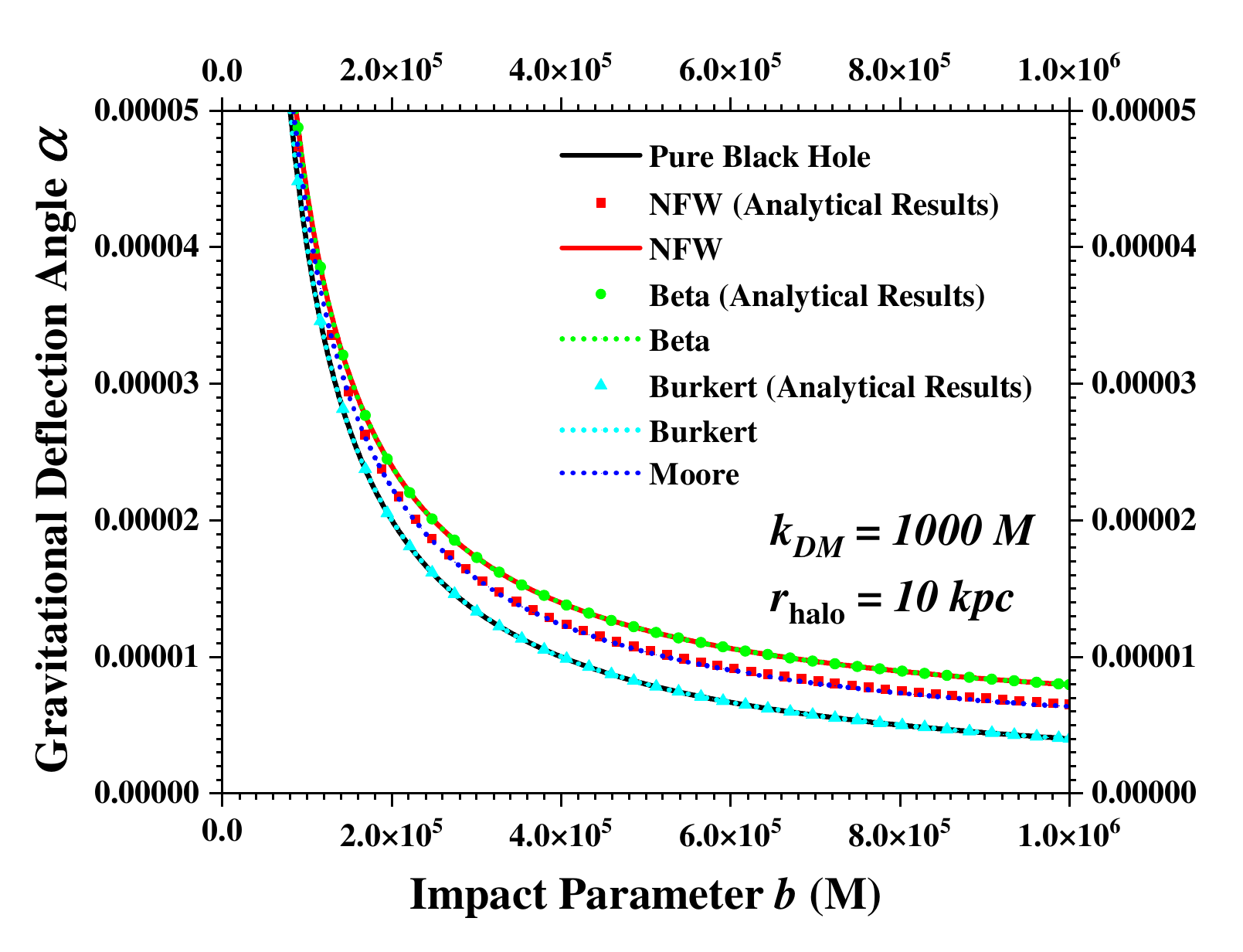}
	\caption{The gravitational deflection angle of light in the gravitational lensing of black hole surrounded by dark matter halos. This figure presents the numerical results calculated using the geodesic approach in equation (\ref{gravitational deflection angle for null geodesic}) for black hole surrounded by NFW, Bata, Burkert, Moore dark matter halos. The dark matter halo scale is selected as $r_{\text{halo}}=10$ kpc, and the dark matter characteristic mass is chosen to be $k_{\text{DM}}=10M$ and $k_{\text{DM}}=1000M$ in the left and right panels, respectively. The gravitational deflection angle of light in gravitational lensing of pure black holes without the presence of dark matter is illustrated using black curves for comparisons. Furthermore, analytical results of gravitational deflection angles obtained in the weak field approximation in equations (\ref{gravitational deflection angle analytical Beta}-\ref{gravitational deflection angle analytical Burkert}) for NFW, Beta, Burkert models are also included. In this figure, the locations of observer and light source are fixed as $r_{\text{O}}=r_{\text{S}}=10^{10} M$. The impact parameter in the photon orbit varies from $b=10^{5} M \sim 0.1\text{pc}$ to $b=10^{6} M \sim \text{pc}$ (for the supermassive black hole mass $M=10^{7} M_{\odot} \sim 10^{7} $ km $\sim 10^{-6}$ pc).}
	\label{figure6}
\end{figure*} 

\begin{figure*}
	\includegraphics[width=0.5\textwidth]{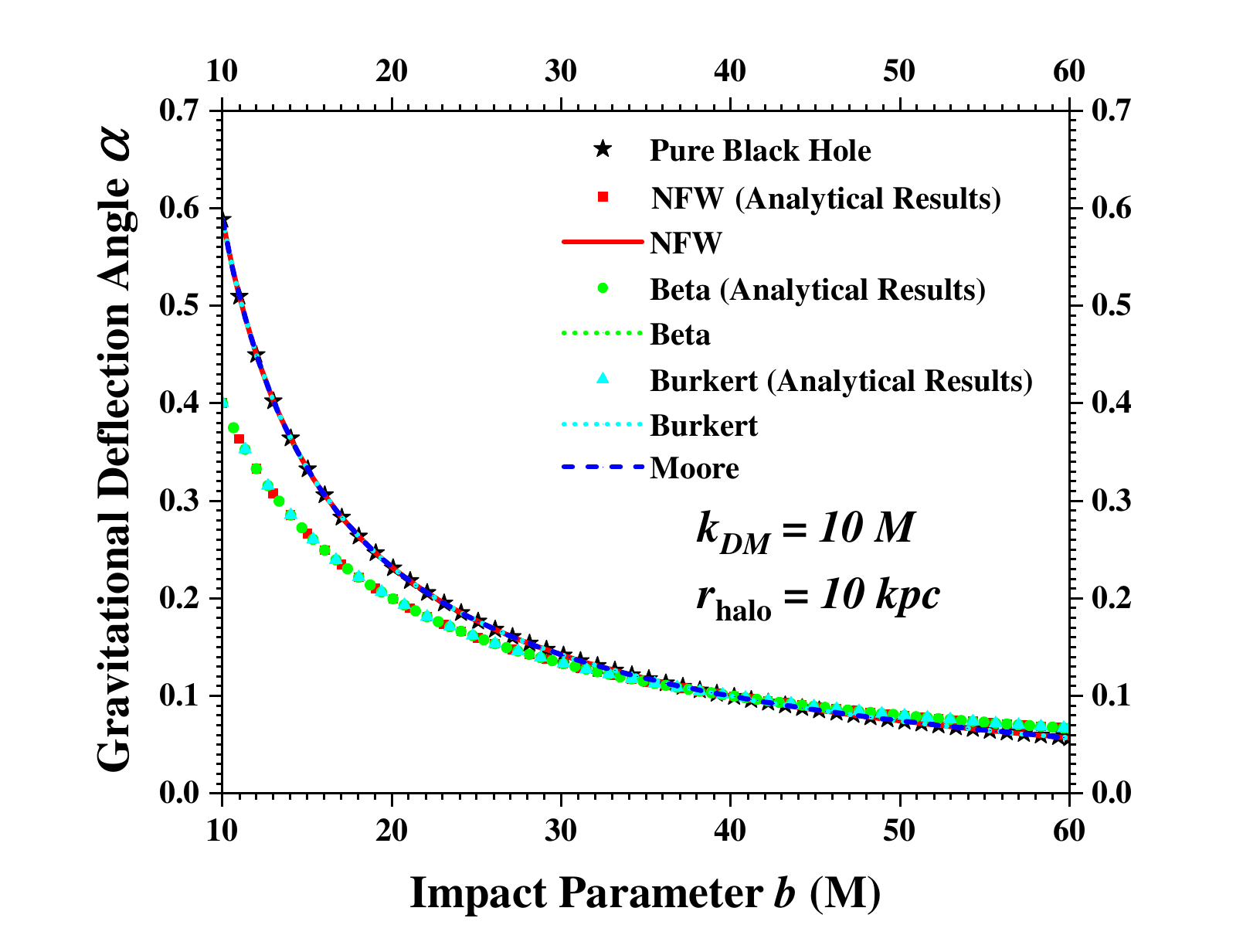}
	\includegraphics[width=0.5\textwidth]{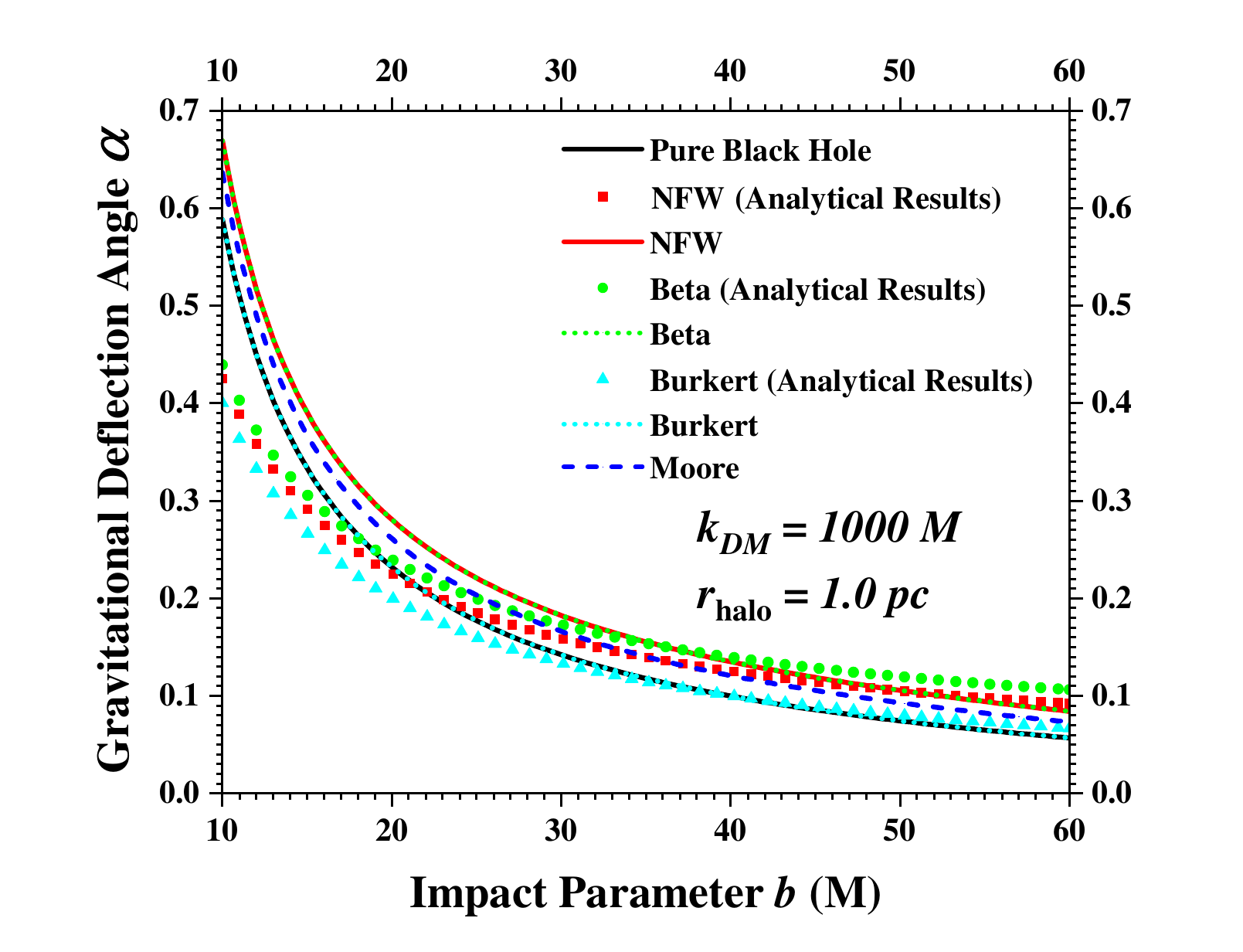}
	\caption{The gravitational deflection angle of light in the gravitational lensing of black hole surrounded by dark matter halos in the strong gravitational field cases. This figure presents the numerical results calculated using the geodesic approach in equation (\ref{gravitational deflection angle for null geodesic}) for black hole surrounded by NFW, Bata, Burkert, Moore dark matter halos. The left panel corresponds to a typical dark matter halo scale $r_{\text{halo}}=10$ kpc and characteristic mass $k_{\text{DM}}=10M$ in galaxies. The right panel gives an extreme parameter choice such that gravitational deflection angles from different halo models can be distinguished, with a larger dark matter mass $k_{\text{DM}}=1000M$ and an ultra small dark matter halo scale $r_{\text{halo}}=1$ pc. The gravitational deflection angle for pure black hole cases without the presence of dark matter, as well as the analytical results on gravitational deflection angle obtained in the weak field approximation in equations (\ref{gravitational deflection angle analytical Beta}-\ref{gravitational deflection angle analytical Burkert}) for NFW, Beta, Burkert models, are also included in this figure for comparisons. To display the strong gravitational field cases, the locations of observer and light source are selected near the central black hole, at $r_{\text{O}}=r_{\text{S}}=100 M$. The impact parameter in the photon orbit varies from $b=10 M$ to $b=60 M$.}
	\label{figure7}
\end{figure*}

The numerical results on gravitational deflection angles in the gravitational lensing of supermassive black holes surrounded by dark matter halos are given in figure \ref{figure6} and figure \ref{figure7}. The figure \ref{figure6} shows the gravitational deflection angles calculated with typical galactic parameters (which are in the weak gravitational field cases), while the figure \ref{figure7} provides the gravitational deflection angles calculated in strong gravitational field cases. Particularly, in the weak gravitational field cases illustrated in figure \ref{figure6}, the analytical results of gravitational deflection angle are also presented for comparisons. The analytical expressions of gravitational deflection angle of light for black hole surrounded by NFW, Beta, Burkert dark matter halos in the weak gravitational field limit (where observer and light source are located at infinity) have been given in recent works \cite{Qiao2023b,Pantig2022} 
\footnote{The analytical result of gravitational deflection angle of light for black hole surrounded by Moore dark matter halo is also obtained in the recent work \cite{Qiao2023b}.  However, the analytical expression has a number of special functions (both the special functions used in mathematics and the special functions defined by authors), which makes it inconvenient to give a comparison in the present study. Thus, we do not plot the analytical result corresponds to the Moore dark matter halo model in figure \ref{figure6}.}
\begin{subequations}
\begin{eqnarray}
	\alpha_{\text{NFW}}  & = & \frac{4M}{b} 
	                           + \frac{8 \pi k_{\text{DM}}}{b} \cdot \ln \bigg( 1+\frac{b}{r_{\text{halo}}} \bigg)
	                           \approx \frac{4M}{b} + \frac{8\pi k_{\text{DM}}}{r_{\text{halo}}} 
	                           \label{gravitational deflection angle analytical NFW}
	                           \\
	\alpha_{\text{Beta}} & = & \frac{4M}{b} 
	                           + \frac{4 \pi k_{\text{DM}}}{r_{\text{halo}}} \bigg( \pi -  \frac{b^{3}}{r_{\text{halo}}^{3}} \bigg) 
	                           \label{gravitational deflection angle analytical Beta}
	                              \\
	\alpha_{\text{Burkert}} & = & \frac{4M}{b} 
	                              + \frac{\pi k_{\text{DM}} b^{3}}{18r_{\text{halo}}^{4}} \cdot \bigg( 5-3\pi+12 \ln \frac{b}{2r_{\text{halo}}} \bigg) .
	                              \label{gravitational deflection angle analytical Burkert}
\end{eqnarray}
\end{subequations}
The analytical results plotted in figure \ref{figure6} are calculated form these expressions. In contrast to the cases of time delay, different dark matter halo models are unable to be distinguished from the gravitational deflection angles for a typical choice of galactic dark matter parameters, with characteristic mass $k_{\text{DM}} = 10M$ and scale length $r_{\text{halo}} = 10$ kpc (at the distance $r_{\text{O}}=r_{\text{S}}=10^{10} M$). The analytical and numerical results obtained within various dark matter halo models do not have obvious differences, and they overlap each other in the left panel of figure \ref{figure6}. To effectively distinguish these dark matter halo models from comparisons of gravitational deflection angles, the mass of dark matter halo should be increased to a sufficiently large value (for a typical dark matter halo scale $r_{\text{halo}} = 10$ kpc in galaxies). The results plotted in the right panel of figure \ref{figure6} indicate that the dark matter mass should be increased to a magnitude of $k_{\text{DM}} = 1000 M$ to make the gravitational deflection angles in various dark matter halo models different \footnote{Unfortunately, in the right panel of figure \ref{figure6}, the numerical results on gravitational deflection angles calculated using NFW and Beta models are overlapped, and the gravitational deflection angles obtained from Burkert model coincide with those for pure black hole cases without dark matter. For the dark matter halo scale $r_{\text{halo}} = 10$ kpc and dark matter mass $k_{\text{DM}} = 1000 M$, it remains not possible to distinguish between NFW model and Beta model (or distinguish between Burkert model and pure black hole cases), from the comparison of gravitational deflection angle data.}. Specifically, when $k_{\text{DM}} = 1000 M$ and $r_{\text{halo}} = 10$ kpc, the NFW and isothermal Beta halo models results in largest deflection angles, while the Burkert model's results do not exhibit obvious differences with the pure black hole's results. Moreover, the analytical and numerical gravitational deflection angles for black hole surrounded by Beta and Burkert halo models exhibit a high level of consistency. However, there are slight deviations between the analytical and numerical results obtained from NFW model (the analytical results correspond to NFW model are much closer to the numerical results for Moore model, as shown in the right panel of figure \ref{figure6}), possibly owning to the lack of higher order contributions in the analytical expressions (\ref{gravitational deflection angle analytical NFW}).

In the strong gravitational cases, the gravitational deflection angles of light in the gravitational lensing of black hole surrounded by dark matter halos are given in figure \ref{figure7}. To make the gravitational field strong, both observer and light source should be located near the central supermassive black hole, so we choose $r_{\text{O}}=r_{\text{S}}=100 M$ in this figure. The left panel of this figure presents results for typical galactic dark matter parameters ($r_{\text{halo}}=10$ kpc and $k_{\text{DM}}=10M$). However, under these parameters, the contributions from dark matter halos to deflection angle are negligible. Numerical and analytical results from different dark matter halo models do not have obvious differences, and they almost identical to the pure black hole cases without the presence of dark matter. Furthermore, the analytical results for NFW, Beta, Burkert dark matter halo models obtained in equations (\ref{gravitational deflection angle analytical Beta}-\ref{gravitational deflection angle analytical Burkert}) all converge to the classical leading order expression $\alpha_{\text{Sch}} = 4M/b$ for Schwarzschild black hole. To efficiently distinguish different dark matter models from gravitational deflection angles, the dark matter halo characteristic mass and length scale should be adjusted to extremely abnormal values, even more extreme than those required for time delay in figure \ref{figure4}. Since a denser distribution of dark matter can significantly enhance their gravitational influences (both on gravitational deflection and time delay), so a larger dark matter mass and a much smaller halo scale are needed to make noticeable differences for gravitational deflection angles obtained from various dark matter models. Particularly, the right panel of figure \ref{figure7} indicates that dark matter halo mass should be around $k_{\text{DM}} \sim 1000 M$, meanwhile the dark matter halo scale should be as small as $r_{\text{halo}} = 10^5 M \sim 1 \text{pc}$ (for a typical supermassive black hole mass $M=10^{7} M_{\odot} \sim 10^{7} $ km $\sim 10^{-6}$ pc), which is too small to realistically represent a dark matter halo in any galaxies 
\footnote{Unfortunately, in the right panel of figure \ref{figure7}, the numerical results on gravitational deflection angles calculated using NFW and Beta models are overlapped, and the gravitational deflection angles obtained from Burkert model are almost identical to those for pure black hole cases without dark matter. So it is still not possible to distinguish between NFW model and Beta model, or distinguish between Burkert model and pure black hole cases, solely from the comparison of gravitational deflection angle data. But one needn't to worry about this point, since a dark matter halo with such a small scale $r_{\text{halo}}=1$ pc is unrealistic in any galaxies. The purpose to present figure \ref{figure7} is to give an example highlighting situations where gravitational deflection angle could be significantly influenced by dark matter halos, rather than comparing them directly with any observational data.}.

\section*{Acknowledgment}

This work is supported by the Scientific Research Program of Chongqing Science and Technology Commission (the Chongqing “zhitongche” program for doctors, Grant No. CSTB2022BSXM-JCX0100), the Natural Science Foundation of Chongqing Municipality (Grant No. CSTB2022NSCQ-MSX0932), the Scientific and Technological Research Program of Chongqing Municipal Education Commission (Grant No. KJQN202201126), and the Research and Innovation Team Cultivation Program of Chongqing University of Technology (Grant No. 2023TDZ007).

\end{document}